\documentclass[10pt]{IEEEtran}
\usepackage{graphicx}
\usepackage{color}
\usepackage{dsfont}
\usepackage{bbm}
\usepackage{amsmath}
\usepackage{amssymb}
\usepackage{float}
\usepackage{psfrag}
\usepackage{mathdots}
\usepackage{arydshln}
\usepackage{verbatim}
\usepackage{cite}
\graphicspath{{figures/}}

\newcommand{\pan}{\mathcal{P}}

\newcommand{\vb}[1]{\mathbf{#1}}
\newcommand{\vbhat}[1]{\vb{\hat #1}}
\newcommand{\numeq}[2]{\begin{equation} #2 \label{#1} \end{equation}}
\newcommand\sups[1]{^{\hbox{\scriptsize{#1}}}}
\newcommand\supt[1]{^{\hbox{\tiny{#1}}}}
\newcommand\subs[1]{_{\hbox{\scriptsize{#1}}}}
\newcommand\subt[1]{_{\hbox{\tiny{#1}}}}
\newcommand{\nn}{\nonumber \\}
\newcommand\supsstar[1]{^{\hbox{\scriptsize{#1}}*}}
\newcommand\suptstar[1]{^{\hbox{\tiny{#1}}*}}
\newcommand{\inp}[2]{ \big\langle #1 , #2 \big\rangle}

\newcommand{\DO}{\partial\mathcal{O}}
\newcommand{\IF}{_{i\text{\scriptsize F}}}
\newcommand{\IFAB}{_{i\text{\scriptsize F}\alpha\beta}}
\newcommand{\IT}{_{i\text{\scriptsize T}}}
\newcommand{\ITAB}{_{i\text{\scriptsize T}\alpha\beta}}
\newcommand{\mc}{\mathcal}
\newcommand{\wt}{\widetilde}
\newcommand{\vbGamma}{\boldsymbol{\Gamma}}

\newcommand{\citeasnoun}[1]{Ref.~\citen{#1}}
\begin{document}

\title{Efficient Computation of Power, Force, and Torque in BEM Scattering Calculations}
\author{M.~T.~Homer~Reid%,
        \thanks{M. T. Homer Reid is with the Department of Mathematics, 
                Massachusetts Institute of Technology.}
%
%        Alejandro W. Rodriguez,%
%       \thanks{Alejandro Rodriguez is with the School of Engineering 
%               and Applied Sciences, Harvard University, and 
%               the Department of Mathematics, Massachusetts Institute 
%               of Technology}
        and Steven~G.~Johnson%
        \thanks{S. G. Johnson is with the Department of Mathematics,
                Massachusetts Institute of Technology.}
       }
\maketitle

%%%%%%%%%%%%%%%%%%%%%%%%%%%%%%%%%%%%%%%%%%%%%%%%%%%%%%%%%%%%%%%%%%%%%%
%%%%%%%%%%%%%%%%%%%%%%%%%%%%%%%%%%%%%%%%%%%%%%%%%%%%%%%%%%%%%%%%%%%%%%
%%%%%%%%%%%%%%%%%%%%%%%%%%%%%%%%%%%%%%%%%%%%%%%%%%%%%%%%%%%%%%%%%%%%%%
%\input{Abstract}
\begin{abstract}
We present concise, computationally efficient formulas for several
quantities of interest---including absorbed and scattered power, 
optical force (radiation pressure), and torque---in scattering calculations
performed using the boundary-element method (BEM) [also known as the 
method of moments (MOM)]. Our formulas compute the quantities of interest 
\textit{directly} from the BEM surface currents with no need ever to 
compute the scattered electromagnetic fields. We derive our new formulas and 
demonstrate their effectiveness by computing power, force, and torque 
in a number of example geometries. Free, open-source software implementations 
of our formulas are available for download online.
\end{abstract}
%%%%%%%%%%%%%%%%%%%%%%%%%%%%%%%%%%%%%%%%%%%%%%%%%%%%%%%%%%%%%%%%%%%%%%
%%%%%%%%%%%%%%%%%%%%%%%%%%%%%%%%%%%%%%%%%%%%%%%%%%%%%%%%%%%%%%%%%%%%%%
%%%%%%%%%%%%%%%%%%%%%%%%%%%%%%%%%%%%%%%%%%%%%%%%%%%%%%%%%%%%%%%%%%%%%%

%%%%%%%%%%%%%%%%%%%%%%%%%%%%%%%%%%%%%%%%%%%%%%%%%%%%%%%%%%%%%%%%%%%%%%
%%%%%%%%%%%%%%%%%%%%%%%%%%%%%%%%%%%%%%%%%%%%%%%%%%%%%%%%%%%%%%%%%%%%%%
%%%%%%%%%%%%%%%%%%%%%%%%%%%%%%%%%%%%%%%%%%%%%%%%%%%%%%%%%%%%%%%%%%%%%%
\section{Introduction}

This paper presents concise new formulas for the 
absorbed and scattered power, force (radiation pressure), 
and torque exerted on material bodies by
incident fields. Our formulas, which are derived 
in the context of the frequency-domain boundary-element 
method [BEM, also known as the method of moments (MOM)],
express powers, forces, and torques (PFTs) 
in terms of vector--matrix--vector products involving
the vectors of BEM surface-current coefficients
and certain sparse matrices describing basis-function
overlaps. We present the derivation of our formulas and 
apply them to a number of example geometries.

The primary goal of electromagnetic scattering
solvers is to compute the electric and magnetic 
fields in a given material geometry illuminated 
by given incident fields or sources. However, in
many cases we are less interested in the fields 
themselves than in certain derived quantities 
obtained from them. For example, in a scattering
problem we may be more interested in the total 
absorbed power or the total scattering cross section
than in the individual fields at particular points in
space~\cite{Eibert2007, Uluisik2008}. Similarly, for 
the problem of a mesoscopic 
structure~\cite{Collett2003,Makarov2004} or 
nanoparticle~\cite{Ashkin1970, Ignatovich2003, Sun2006}
illuminated by a laser, we may seek the force or torque 
exerted on the particle by the incident-field sources.

Of course, derived quantities such as power, force, 
and torque may always be computed \textit{indirectly} 
in any scattering formalism by first computing the
scattered fields and then post-processing: powers and 
forces/torques are obtained respectively by integrating 
the Poynting vector (PV) and the Maxwell stress tensor 
(MST) over a bounding surface surrounding the object in 
question, with values of the PV and MST at each surface 
point computed using field components obtained 
from the scattering solver. However, in practice
the integration may be numerically badly behaved
due to large cancellations from different regions
of the bounding surface, requiring large numbers
of cubature points to obtain accurate results. Moreover, 
in scattering formalisms such as the BEM---where, unlike
other techniques such as the finite-difference and 
finite-element methods, the fields are not computed 
directly as part of the solution to the problem---each 
field evaluation required to evaluate the PV or MST
integrals by numerical cubature costs extra work.

To date a variety a methods have been used to compute 
PFTs for compact bodies. For bodies of highly 
symmetric shapes such as spheres or cylinders, exact 
results are available in analytical form (such as Mie 
theory for spheres)~\cite{BohrenHuffman1983, Harrington1961}, 
and these closed-form solutions may be used to compute PFTs 
for such bodies~\cite{Barton1988, Almaas1995, Kemp2006A, Kemp2006B},
or extended to the case of spheroidal bodies~\cite{Xu2007}.
For bodies small compared to the incident wavelength 
(``Rayleigh particles''), PFTs may be computed 
using quasistatic 
approximations~\cite{Leung1986,Ng2000,Going2008,Chen2009,Chen2012}, 
while in the opposite regime (bodies large compared to the 
incident wavelength) the geometrical-optics approximation
is available~\cite{Gussgard1992, Gauthier1995, Gu2000}.
Radiation forces for nonspherical bodies
have also been computed using quasi-analytical 
approaches (numerical methods based on analytical
solutions for special geometries), 
including T-matrix methods~\cite{Nieminen2007} and 
discrete dipole approximations~\cite{Ling2010},
Among fully numerical techniques, compact-object 
PFTs have been computed using finite-difference 
time-domain (FDTD) 
methods~\cite{Taflove1989, Collett2003, Sun2006, Dewan2011, Cao2012}
and finite-element methods~\cite{McNeil2001, Barth2006, Huda2013}.

BEM techniques have been used by several authors
to characterize \textit{local} power
absorption~\cite{Chuang1994, Chuang1997, Chen1998, 
Bottauscio2012}, and were used in~\citeasnoun{Kern2009}
to compute total cross sections for scattering 
and absorption by a compact body. \citeasnoun{Makarov2004}
employed a BEM technique to investigate the force and 
torque on a mesoscopic body illuminated by a laser beam,
while \citeasnoun{Xiao2008} used the BEM to compute
radiation forces on two-dimensional structures.
In all of these BEM studies, the standard BEM approach
was applied to solve for surface currents, after
which the surface currents were used to compute
scattered fields. The total absorbed/scattered
power~\cite{Kern2009} or force~\cite{Makarov2004,Xiao2008}
was obtained by numerically evaluating a surface 
integral over the body or over a bounding 
surface.

An alternative BEM approach to the computation of 
absorbed power was suggested by \citeasnoun{Kern2010}, 
which noted that the power absorbed by a compact body 
may be obtained directly from the surface currents,
obviating the intermediate step of computing
scattered fields. (This observation seems to have
been mentioned as something of a passing curiosity 
in \citeasnoun{Kern2010}; although the authors
note that the absorption may be computed directly 
from the surface currents, computational details 
are omitted, and thereafter the authors revert to 
the usual practice of obtaining absorption and
other quantities via the intermediate device of 
computing scattered fields.)

The objective of this paper is to extend the observation
of \citeasnoun{Kern2010} by demonstrating that, in the BEM 
framework, powers, forces, and torques 
(PFTs) may be computed \textit{directly} from the surface 
currents that are the primary output of the BEM solver.
This approach bypasses unwieldy numerical cubatures of the PV
and MST, yielding concise formulas expressing PFTs in terms 
of vector--matrix--vector products (VMVPs). The vectors that enter
these expressions are just the RHS vector and the solution 
vector of the linear BEM system, while the matrices describe
various overlaps of the surface-current basis functions and 
have the virtue of being highly \textit{sparse} for localized 
basis functions (including the commonly used Rao-Wilton-Glisson 
(RWG) functions~\cite{RWG}). The upshot is that we can 
compute PFTs immediately upon assembling and solving the 
linear BEM system, with no need ever to compute a scattered
field. 

In addition to the theoretical compactness and 
practical convenience of our formulas, their computational
impact is to render essentially negligible the cost of 
evaluating PFTs beyond the base cost of solving 
the linear BEM system. More specifically, for a BEM 
scattering problem with surface currents expanded in 
a basis of $N$ localized functions (such as RWG functions),
our PFT algorithms entail computational cost scaling like 
$O(N)$ with a small constant prefactor
once the linear BEM system has been solved.
%^an insignificant burden compared to the 
%$O(N^3)$ (for dense-direct solvers) or 
%$O(N\log N)\sim O(N^2\log N)$ (for fast solvers)
%cost of solving the BEM system in the first place.
Of course, in geometries involving only a single
body---or multiple bodies in which we require only
aggregate PFTs for the collection of all bodies---PFTs
may generally be computed in $O(N)$ time by 
evaluating numerical cubatures of the PV or MST over 
distant bounding surfaces. In this case, the distance
of the bounding surface enables the use of numerical 
cubature with a number of evaluation points that is
independent of the fineness of the mesh discretization;
the computational cost would then be just that of 
evaluating the scattered fields at each cubature 
point, an $O(N)$ procedure in BEM. 
However, in cases involving closely-spaced bodies
in which we require individual PFTs for each
body---such as the two closely-spaced nanoparticles
of Figure \ref{ChiralParticlePairFigure}---numerical 
cubature of the PV or MST over a bounding
surface surrounding just one body would presumably require
large numbers of cubature points in the region between
the two bodies, and indeed the number of cubature
points may grow as the mesh is refined, which would
increase the computational cost scaling to $O(N^2).$
Moreover, asymptotic considerations aside, in 
practice the computation of PFTs via surface integration
of the PV or MST entails the cumbersome steps of defining 
a bounding surface and implementing a multidimensional
numerical cubature with an unwieldy integrand.
These complications are entirely bypassed by the concise 
VMVP formulas we present here.

Our VMVP formulas for PFTs are closely related to
certain determinant and trace formulas for Casimir
forces and radiative heat-transfer rates arising
in the ``fluctuating-surface-current'' (FSC)
approach to fluctuation
phenomena~\cite{Reid2012A, Reid2012B, Rodriguez2013}.
The FSC approach addresses problems involving
thermal and quantum-mechanical fluctuations in
the electromagnetic field; the present paper may
be viewed as an exploration of the applications
of FSC ideas to classical scattering calculations.
 
The remainder of this paper is organized as follows. In 
Section \ref{PowerSection} we derive VMVP formulas for the
power absorbed by and scattered from objects irradiated
by external fields. In Section \ref{ForceTorqueSection} we
derive VMVP formulas for the force and torque exerted on 
objects by external fields. In Section \ref{ExamplesSection}, 
we first validate our formulas by using them to reproduce 
known results for spheres, where Mie theory may be used to 
compute PFTs analytically. Then we illustrate the usefulness 
of our formulas by applying them to bodies of complicated 
shapes for which other methods of computing PFTs would be 
unwieldy. Our conclusions are presented in Section 
\ref{ConclusionsSection}.
Appendix \ref{AlternativeScatteredPowerAppendix} presents
an alternative physical interpretation of the power 
formulas derived in Section \ref{PowerSection}, which
has the added benefit of leading to an alternative VMVP
formula for the scattered power.

The formulas derived in Sections \ref{PowerSection} and 
\ref{ForceTorqueSection} involve matrices whose entries
are overlap integrals involving the basis functions used 
to represent surface currents in the BEM solver. These 
formulas are fully general, in the sense that they refer 
to no specific choice of basis functions. In Appendix 
\ref{RWGAppendix}, we specialize to the particular
choice of RWG basis functions~\cite{RWG} and present 
explicit formulas for the elements of the power, 
force, and torque matrices in this case.
%%%%%%%%%%%%%%%%%%%%%%%%%%%%%%%%%%%%%%%%%%%%%%%%%%%%%%%%%%%%%%%%%%%%%%
%%%%%%%%%%%%%%%%%%%%%%%%%%%%%%%%%%%%%%%%%%%%%%%%%%%%%%%%%%%%%%%%%%%%%%
%%%%%%%%%%%%%%%%%%%%%%%%%%%%%%%%%%%%%%%%%%%%%%%%%%%%%%%%%%%%%%%%%%%%%%

%%%%%%%%%%%%%%%%%%%%%%%%%%%%%%%%%%%%%%%%%%%%%%%%%%%%%%%%%%%%%%%%%%%%%%
%%%%%%%%%%%%%%%%%%%%%%%%%%%%%%%%%%%%%%%%%%%%%%%%%%%%%%%%%%%%%%%%%%%%%%
%%%%%%%%%%%%%%%%%%%%%%%%%%%%%%%%%%%%%%%%%%%%%%%%%%%%%%%%%%%%%%%%%%%%%%
\section{Absorbed, Scattered, and Total Power}
\label{PowerSection}

%%%%%%%%%%%%%%%%%%%%%%%%%%%%%%%%%%%%%%%%%%%%%%%%%%%%%%%%%%%%%%%%%%%%%%
%%%%%%%%%%%%%%%%%%%%%%%%%%%%%%%%%%%%%%%%%%%%%%%%%%%%%%%%%%%%%%%%%%%%%%
%%%%%%%%%%%%%%%%%%%%%%%%%%%%%%%%%%%%%%%%%%%%%%%%%%%%%%%%%%%%%%%%%%%%%%
\subsection{Preliminaries; Notation}

We consider a scattering geometry 
%like that depicted in Figure \ref{CompactFigure} 
consisting of a compact material body 
$\mathcal{O}$---possibly in the presence
of other compact or extended material bodies---illuminated 
by known incident fields. The object(s) may be 
embedded in a dielectric medium, but if so we 
assume this medium to be \textit{lossless}.
Let $\DO$ be the 
surface of $\mathcal{O}$ and $\vbhat{n}(\vb x)$ the 
outward-pointing normal vector at a point $\vb x$
on $\DO$.
We work in the frequency domain with all fields and
currents understood to have time dependence 
$\propto e^{-i\omega t}$.

In BEM formulations for dielectric 
scatterers---such as the PMCHW~\cite{Medgyesi1994} or
N-M\"uller~\cite{Taskinen2005} formulations---the quantities 
of immediate interest are the electric 
and magnetic \textit{surface currents} 
$\vb K(\vb x)$ and $\vb N(\vb x)$, which are defined,
for points $\vb x$ on the surfaces of material bodies, 
in terms of the tangential components of the 
total magnetic and electric fields:
\numeq{KNDefinition}
{  \vb K(\vb x) \equiv \vbhat{n}(\vb x) \times \vb H(\vb x), 
   \qquad
   \vb N(\vb x) \equiv -\vbhat{n}(\vb x) \times \vb E(\vb x).
}
For numerical computations, $\vb K$ and $\vb N$ are 
approximated as expansions in some discrete set of 
tangential-vector-valued basis functions $\{\vb b_\alpha\}$ 
defined on the object surfaces:
\numeq{KNExpansion}
{ \vb K(\vb x) \approx \sum_\alpha k_\alpha \vb b_\alpha(\vb x), \qquad 
  \vb N(\vb x) \approx \sum_\alpha n_\alpha \vb b_\alpha(\vb x)
} 
One commonly chosen set of basis functions are the
RWG functions~\cite{RWG}, and later in this paper we will 
present specific results and formulas for RWG basis functions, 
but the techniques we introduce here are not restricted to any 
particular choice of surface-current expansion functions.

To find the surface currents induced by given
incident fields $\{\vb E\sups{inc}, \vb H\sups{inc}\},$
the $\{k_\alpha, n_\alpha\}$ coefficients are 
obtained by solving a linear system of the form
\numeq{BEMSystem}
{ 
  \left(\begin{array}{cc} 
    \vb M\supt{EE} & \vb M\supt{EM} \\ 
    \vb M\supt{ME} & \vb M\supt{MM} \\ 
  \end{array}\right)
  \left(\begin{array}{c} \vb k \\ \vb n \end{array} \right)
  =
  -
  \left(\begin{array}{c} \vb v\sups{E} \\ \vb v\sups{H} \end{array}\right)
}
where the elements of the RHS vector involve the 
incident fields; for example, in the PMCHW formulation
the RHS elements are just the inner products of 
the incident fields with the basis functions:
\begin{subequations}
\begin{align}
  v\sups{E}_\alpha 
   &= \int_{\sup \vb b_{\alpha}} \vb b^*_\alpha(\vb x) \cdot \vb E\sups{inc}(\vb x)\,d\vb x
   \nonumber\\
   &\equiv \inp{\vb b_\alpha}{\vb E\sups{inc}}\\
 \intertext{and similarly}
  v\sups{H}_\alpha &= \inp{\vb b_\alpha}{\vb H\sups{inc}}.
\end{align}
\label{RHSDefinition}
\end{subequations}
For convenience in what follows we will write equation (\ref{BEMSystem})
in the form 
\numeq{BEMSystem2}{\vb M \vb c = -\vb v}
with 
$$\vb c=\left(\begin{array}{c} \vb k \\ \vb n \end{array}\right)
  \qquad
  \vb v=\left(\begin{array}{c} \vb v\sups{E} \\ \vb v\sups{H} \end{array}\right).
$$

\textit{PEC bodies.} The preceding discussion applies to the case of 
dielectric bodies.  For perfectly electrically conducting (PEC) bodies, 
formulations such as the electric-field integral equation 
(EFIE)~\cite{Harrington93} involve equations similar to the above 
but with only electric surface-current unknowns; the magnetic surface 
current $\vb N$ vanishes in this case. All formulas given in this paper 
apply equally well to PEC bodies and may be used as-is by simply setting
to zero all magnetic surface-current unknowns and all magnetic 
RHS vector elements, $n_\alpha=v\sups{H}_\alpha=0$ for all $\alpha$.

%%%%%%%%%%%%%%%%%%%%%%%%%%%%%%%%%%%%%%%%%%%%%%%%%%%%%%%%%%%%%%%%%%%%%%
%%%%%%%%%%%%%%%%%%%%%%%%%%%%%%%%%%%%%%%%%%%%%%%%%%%%%%%%%%%%%%%%%%%%%%
%%%%%%%%%%%%%%%%%%%%%%%%%%%%%%%%%%%%%%%%%%%%%%%%%%%%%%%%%%%%%%%%%%%%%%
\subsection{Absorbed Power}

The power flowing into the body
is the integral of the inward-directed normal component of 
the total Poynting vector over its surface:
%====================================================================%
\begin{align}
 P\sups{abs} 
&= - \int_{\DO} \vb P\sups{tot}(\vb x) \cdot \vbhat{n}(\vb x) \, dA 
\label{AbsorbedPower}
\end{align}
%====================================================================%
where the minus sign arises because by convention we define $\vbhat{n}$ 
to be the outward-directed surface normal.

At a point $\vb x$ on $\DO$, the (outward-directed) normal 
component of the total Poynting vector is 
%====================================================================%
\begin{align}
 \vb P\sups{tot}(\vb x) \cdot \vbhat{n}(\vb x)
  &=\frac{1}{2}\text{Re} 
    \bigg\{ \vbhat{n}(\vb x) \cdot 
            \Big[\vb E\supsstar{tot}(\vb x) \times \vb H\sups{tot}(\vb x)
            \Big]
    \bigg\}
\\
%--------------------------------------------------------------------%
\intertext{Using the fact that $|\vbhat{n}|=1$, we may rewrite this 
           in the form (temporarily suppressing $\vb x$ arguments)}
  &=-\frac{1}{2}\text{Re} 
    \Bigg\{ \big(\vbhat{n}\times \vb H\supsstar{tot}\big)
            \cdot
            \big(\vbhat{n} \times \big(-\vbhat{n} \times \vb E\sups{tot}\big)
            \big)
    \Bigg\}.
\nonumber
%--------------------------------------------------------------------%
\intertext{But the quantities in parentheses here are just the effective 
           electric and magnetic surface currents (\ref{KNDefinition}),
           so we find simply}
%--------------------------------------------------------------------%
\vb P\sups{tot}(\vb x) \cdot \vbhat{n}(\vb x)
  &=-\frac{1}{2}\text{Re}
     \bigg\{ \vb K^* \cdot \Big[\vbhat{n} \times \vb N \Big] \bigg\}.
\label{PTotDotN}
\end{align}
%====================================================================%
%
% some quick matlab code that verifies the above:
%  E=rand(3,1) + sqrt(-1)*rand(3,1);
%  EStar=conj(E);
%  H=rand(3,1) + sqrt(-1)*rand(3,1);
%  nHat=rand(3,1); 
%  nHat=nHat/norm(nHat);
%  K=cross(nHat,H);
%  KStar=conj(K);
%  N=-cross(nHat,E);
%  EStarxH=cross(conj(E),H);
%  P1 = 0.5*real( nHat' * EStarxH )
%  P2 = -0.5*real( KStar.' * cross(nHat, N) );
[As noted in the Introduction, this approach was
anticipated by \citeasnoun{Kern2010}; equation 
(\ref{PTotDotN}) is equivalent to equation (10) 
in that work.]
Inserting (\ref{PTotDotN}) into (\ref{AbsorbedPower}), we have
%====================================================================%
\begin{align}
 P\sups{abs}
&=\frac{1}{2}\text{Re }\int_{\DO}
  \bigg\{ \vb K^*(\vb x) \cdot 
          \Big[\vbhat{n} \times \vb N(\vb x) \Big] 
  \bigg\} d\vb x.
\nonumber\\
%--------------------------------------------------------------------%
\intertext{
Now inserting the surface-current expansions (\ref{KNExpansion}),
we find}
%--------------------------------------------------------------------%
&=\frac{1}{2}\text{Re }\sum_{\alpha\beta}  k^*_\alpha n_\beta
  \int  \vb f^*_\alpha(\vb x) \cdot  
         \Big[\vbhat{n} \times \vb f_\beta(\vb x) \Big]
         d\vb x
\nonumber\\
&=\frac{1}{2}\text{Re }\sum_{\alpha\beta} 
    k^*_\alpha O^{(\times)}_{\alpha\beta} n_\beta
\label{PAbsFormula0}
\end{align}
%====================================================================% where the $\alpha,\beta$ sum runs only over the indices of
where the sum is over all
basis functions defined on the surface of $\mathcal{O}$, and 
where we have introduced the 
``crossed overlap matrix,'' $\vb O^{(\times)},$
whose elements describe a certain type of overlap between 
basis functions:
\begin{align*}
O^{(\times)}_{\alpha\beta}
=\int \vb f^*_\alpha(\vb x) \cdot 
       \Big[\vbhat{n} \times \vb f_\beta(\vb x) \Big]
       d\vb x.
\end{align*}
%====================================================================%
If we work with a basis of localized functions, then this 
matrix is sparse; for RWG basis functions, it contains 
precisely 4 nonzero entries per row, which may be computed
in closed form (Appendix \ref{RWGAppendix}).

For PEC bodies we have $n_\beta=0$ for all $\beta$ 
and equation (\ref{PAbsFormula0}) vanishes, corresponding to 
the inability of a perfect conductor to absorb power from 
incident-field sources.

If we denote the surface-current vector in 
(\ref{BEMSystem}) by $\vb c={\vb k \choose \vb n}$,
then (\ref{PAbsFormula0}) takes the form
\numeq{PAbsFormula}
{\boxed{P\sups{abs} = \frac{1}{4} \vb c^\dagger \vb O\subs{P}\vb c}}
where $\vb O\subs{P}$, the ``power overlap matrix,''
has the block form
\numeq{PowerBlockMatrix}
{\vb O\subs{P}=\left(\begin{array}{cc}
   0                & \vb O^{(\times)} \\ 
  -\vb O^{(\times)} & 0 \\ 
   \end{array}\right).
}
(The minus sign in the lower-left matrix block arises
because $O_{\beta\alpha}^{(\times)}=-O_{\alpha\beta}^{(\times)}$.)
Because (for localized basis functions) $\vb O_P$ is sparse,
the memory and CPU-time cost of evaluating equation
(\ref{PowerBlockMatrix}) scales like $O(N)$, where $N$
is the number of basis functions.

Equation (\ref{PAbsFormula}) gives the the total net power 
flow through the surface $\DO$. If object 
$\mathcal{O}$ is a solid body, this is just
the total power absorbed in $\mathcal{O}$. If $\mathcal{O}$ 
contains one or more nested inclusions, then 
(\ref{PAbsFormula}) gives the total power absorbed by 
$\mathcal{O}$ and all bodies contained within it.
In this case, the power absorbed by the individual 
subobjects may be computed by applying (\ref{PAbsFormula})
to the subsurfaces bounding those objects.

Although we have here derived (\ref{PAbsFormula}) by
considering the surface integral of the Poynting flux,
it is possible to arrive at the \textit{same} formula 
from alternative physical considerations emphasizing 
the work done by fields on currents. This complementary
viewpoint is presented in 
Appendix \ref{AlternativeScatteredPowerAppendix}.

%%%%%%%%%%%%%%%%%%%%%%%%%%%%%%%%%%%%%%%%%%%%%%%%%%%%%%%%%%%%%%%%%%%%%%
%%%%%%%%%%%%%%%%%%%%%%%%%%%%%%%%%%%%%%%%%%%%%%%%%%%%%%%%%%%%%%%%%%%%%%
%%%%%%%%%%%%%%%%%%%%%%%%%%%%%%%%%%%%%%%%%%%%%%%%%%%%%%%%%%%%%%%%%%%%%%
\subsection{Scattered Power}

The power scattered from $\mathcal{O}$ is the integral of the
outward-directed normal component of the \textit{scattered} 
Poynting vector over $\DO$:
%====================================================================%
\begin{align}
P\sups{scat} 
&= + \int_{\DO} \vb P\sups{scat}(\vb x) \cdot \vbhat{n}(\vb x) \, dA 
\label{ScatteredPower}
\end{align}
%====================================================================%
where the scattered Poynting vector is the 
Poynting vector as computed using only the scattered fields.
In analogy to equation (\ref{PTotDotN}), we write
%====================================================================%
\begin{subequations}
\begin{align}
&\vb P\sups{scat}(\vb x) \cdot \vbhat{n}(\vb x)
  =\frac{1}{2}\text{Re } 
    \bigg\{ \vbhat{n}(\vb x) \cdot
            \Big[ \vb E\supsstar{scat}(\vb x) \times \vb H\sups{scat}(\vb x)
            \Big]
    \bigg\}
\nonumber
\intertext{Noting that scattered fields are the differences between
total and incident fields and again suppressing $\vb x$ arguments, we
find}
%--------------------------------------------------------------------%
  &=\frac{1}{2}\text{Re }
    \bigg\{ \vbhat{n}\cdot
            \Big[ \big(\vb E\supsstar{tot} - \vb E\supsstar{inc}\big)
                   \times
                  \big(\vb H\sups{tot} - \vb H\sups{inc}\big)
            \Big]
    \bigg\}
\nonumber
\intertext{which we write as a sum of three terms:}
  &= \hspace{0.18in} \vb P\sups{tot} \cdot \vbhat{n}
\\[3pt]
  & \hspace{0.2in}
     +\frac{1}{2}\text{Re } 
        \bigg\{ \vbhat{n}\cdot \Big[ \vb E\supsstar{inc} \times \vb H\sups{inc} \Big]
        \bigg\}
\\[3pt]
  & \hspace{0.2in}
     -\frac{1}{2}\text{Re }
      \bigg\{ \vbhat{n}\cdot 
              \Big[ \vb E\supsstar{inc} \times \vb H\sups{tot} \Big]
             \, + \, 
             \vbhat{n}\cdot 
              \Big[ \vb E\supsstar{tot} \times \vb H\sups{inc} \Big]
      \bigg\}.
\end{align}
\label{ThreeTerms}
\end{subequations}
\paragraph{First term}
The first term here (\ref{ThreeTerms}a) is the normal component of the 
total Poynting vector, as considered in the previous section; the surface 
integral of this
term yields $-P\sups{abs}.$

\paragraph{Second term}
The second term (\ref{ThreeTerms}b) is the normal Poynting flux
due to the incident field sources alone. The surface integral 
of this term over $\DO$ yields the net power delivered
by the incident field to the volume inside $\DO$
in the absence of $\mathcal{O}$ (and any other material bodies
that may be present in the scattering problem).
But, assuming the ambient medium is lossless,
in the absence of $\mathcal{O}$ there is nowhere
for this power to go; any power that flows into $\DO$
must flow back out, because there is nothing to absorb it.
Hence this term vanishes.

\paragraph{Third Term}

The third term (\ref{ThreeTerms}c) is 
%====================================================================%
\begin{align*}
&\hspace{-0.3in}
 -\frac{1}{2}\text{Re }
      \bigg\{ \vbhat{n}\cdot 
              \Big[ \vb E\supsstar{inc} \times \vb H\sups{tot} \Big]
             \, + \, 
             \vbhat{n}\cdot 
              \Big[ \vb E\supsstar{tot} \times \vb H\sups{inc} \Big]
      \bigg\}
\\
&=
 +\frac{1}{2}\text{Re }
      \bigg\{ \vb E\supsstar{inc} \cdot \Big[\vbhat{n}\times \vb H\sups{tot}\Big]
              \, + \, 
              \vb H\supsstar{inc} \cdot \Big[ -\vbhat{n}\times \vb E\sups{tot}\Big]
      \bigg\}
\\
&=
 +\frac{1}{2}\text{Re }
      \bigg\{ \vb E\supsstar{inc} \cdot \vb K \, + \, 
              \vb H\supsstar{inc} \cdot \vb N
      \bigg\}
\intertext{Insert the surface-current expansions (\ref{KNExpansion}):}
&=
 +\frac{1}{2}\text{Re }\sum_{\alpha}
      \bigg\{ k_\alpha \vb E\supsstar{inc} \cdot \vb f_\alpha 
             \, 
              + n_\alpha \vb H\supsstar{inc} \cdot \vb f_\alpha
      \bigg\}
\intertext{The surface integral of this is }
\int_{\mathcal S} \left\{ \right\} 
&= \frac{1}{2} \text{Re }\sum_{\alpha} 
   \bigg\{ k_\alpha 
           \underbrace{\Big[\int \vb E\supsstar{inc}(\vb x) \cdot \vb f(\vb x) d\vb x\Big]}
                     _{v\suptstar{E}_\alpha}
 \\
&\qquad\qquad\qquad
          +n_\alpha
           \underbrace{\Big[\int \vb H\supsstar{inc}(\vb x) \cdot \vb f(\vb x) d\vb x\Big]}
                     _{v\suptstar{H}_\alpha}
   \bigg\}
\\[5pt]
&= \frac{1}{2}\text{Re } \sum_{\alpha} \Big[v\supsstar{E}_\alpha k_\alpha
                                           +v\supsstar{H}_\alpha n_\alpha \Big]
\end{align*}
where $\{ v\sups{E}_\alpha, v\sups{H}_\alpha\}$ are the 
elements of the RHS vector of the BEM system (\ref{RHSDefinition}).

Combining the three terms of (\ref{ThreeTerms}), the scattered power 
from a compact object is 
\numeq{PScatFormula}
{ \boxed{P\sups{scat} = -P\sups{abs} + \frac{1}{2}\,\text{Re}\,\,\vb v^\dagger \vb c }}
where $\vb c$ and $\vb v$ are respectively the vector of 
surface-current expansion coefficients and the RHS vector in 
(\ref{BEMSystem}).
For PEC bodies, the first term in (\ref{PScatFormula})
vanishes (as discussed above) but the second term does
not. For PEC and non-PEC bodies alike, equation
(\ref{PScatFormula}), like equation (\ref{PAbsFormula}), 
has computational cost scaling like $O(N).$

In cases where scattering is small compared to absorption,
equation (\ref{PScatFormula}) may yield inaccurate results
for finite mesh discretizations; indeed, if 
$P\sups{scat}\ll P\sups{abs}$ then there is a large 
cancellation error in (\ref{PScatFormula}), making the
expression ill-conditioned~\cite{TrefethenBau1997} and 
amplifying numerical errors.
In such cases, the 
alternative physical picture discussed in Appendix 
\ref{AlternativeScatteredPowerAppendix} leads to an
alternative VMVP expression for the scattered power
[equation (\ref{AlternativePScatFormula})] which
avoids this difficulty at the cost of increased 
computational expense.

%%%%%%%%%%%%%%%%%%%%%%%%%%%%%%%%%%%%%%%%%%%%%%%%%%%%%%%%%%%%%%%%%%%%%%
%%%%%%%%%%%%%%%%%%%%%%%%%%%%%%%%%%%%%%%%%%%%%%%%%%%%%%%%%%%%%%%%%%%%%%
%%%%%%%%%%%%%%%%%%%%%%%%%%%%%%%%%%%%%%%%%%%%%%%%%%%%%%%%%%%%%%%%%%%%%%
\subsection{Total Power}

The sum of the absorbed and scattered powers is the total power 
removed from the incident field (sometimes known as the 
\textit{extinction}). From (\ref{PScatFormula}) we see that 
this is simply
%====================================================================%
\numeq{PTotFormula}
{\boxed{P\sups{tot}
        = P\sups{scat} +P\sups{abs}
        = \frac{1}{2}\,\text{Re}\,\,\vb v^\dagger \vb c.
       }}
%====================================================================%
We might alternatively rewrite (\ref{PTotFormula}) by substituting
for $\vb v$ using (\ref{BEMSystem2}):
\numeq{PTotFormula2}
{P\sups{tot}
 = -\frac{1}{2}\,\text{Re}\,\,\vb c^\dagger \vb M \vb c
}
where $\vb M$ is the full BEM matrix.
Because (in the PMCHW formulation) $\vb M$ is known
to be negative semidefinite~\cite{Rodriguez2013},
the RHS of equation (\ref{PTotFormula2}) is guaranteed
to be nonnegative, as expected on physical grounds.

\subsection{Relation to the optical theorem} Equation (\ref{PTotFormula})
may be understood in another way by appealing to the optical
theorem. The amplitude for scattering into a given direction
with a given polarization may be obtained~\cite{Jackson1999} by
projecting the surface currents onto the fields 
of a plane wave traveling in that direction with that 
polarization:
\numeq{Fk}
{  F(\vbhat{k}) 
   = 
   \frac{iZk}{4\pi}\oint_{\partial \mc O}
   \left\{  \vb E\suptstar{PW}\cdot \vb K 
           +\vb H\suptstar{PW}\cdot \vb N 
   \right\}dA
}
Here 
$\{\vb E, \vb H\}\supt{PW}=
 \{\vb E_0,\vbhat{k}\times \vb E_0/Z\}
 e^{i\vb{k} \cdot \vb x}$ 
are the fields of a plane wave with 
polarization vector $\vb E_0$ and 
propagation vector $\vb{k}$, $Z$
and $k=\omega/c$ are the impedance and wavenumber in the
exterior medium, $\vbhat{k}=\vb k/k$, and $F(\vbhat{k})$ 
is defined such that the $\vb E_0$-polarized component of 
the scattered field asymptotically approaches
$\vb E\sups{scat}(\vb x) \to \vb E_0 \frac{e^{ikr}}{r}F(\vbhat{k}).$
Inserting the surface-current expansions (\ref{KNExpansion})
and proceeding as above, we have 
%====================================================================%
\numeq{FkFromVC}
{
 F(\vbhat{k}) = \frac{iZk}{4\pi} \vb v^{\text{\tiny{PW}}\dagger} \vb c
}
%====================================================================%
where $\vb v^{\text{\tiny PW}}$ is the vector of projections
of $\{\vb E,\vb H\}\supt{PW}$ onto the basis functions, similar
to (\ref{RHSDefinition}).
For a plane-wave incident field, the \textit{forward} scattering
amplitude is obtained by taking the fields in (\ref{Fk}) to be
the same as the incident fields, 
$\{\vb E,\vb H\}\supt{PW}=\{\vb E,\vb H\}\sups{inc}$, in which case 
$\vb v$ in (\ref{FkFromVC}) becomes just the RHS vector in 
(\ref{BEMSystem}),
%====================================================================%
$$ F\big(\vbhat{k}=\vbhat{k}\sups{inc}\big)
   = \frac{iZk}{4\pi} \vb v^\dagger \vb c.
$$
%====================================================================%
On the other hand, the optical theorem~\cite{Jackson1999}
relates the forward scattering amplitude to the total power removed 
from a unit-strength plane wave according to
\begin{align*}
 \vb P\sups{tot}
& =\frac{2\pi}{kZ}\,
   \text{Im }\Big[F\big(\vbhat{k}=\vbhat{k}\sups{inc}\big)\Big].
\intertext{Inserting (\ref{FkFromVC}), this reads} 
&=\frac{1}{2}\text{Im }\Big[ i \vb v^\dagger \vb c\Big]
\\
&=\frac{1}{2}\text{Re } \vb v^\dagger \vb c
\end{align*}
in agreement with (\ref{PTotFormula}).

In passing, we note that equation (\ref{FkFromVC}) provides 
an $O(N)$ algorithm for computing the scattering amplitude in 
arbitrary directions: One simply constructs the RHS vector 
corresponding to an incident plane wave of the desired 
direction and polarization, then computes the dot product 
of this vector with the surface-current vector to obtain 
the amplitude for scattering in that direction with that 
polarization. Of course, since the scattering amplitude 
involves only far fields, it may be computed numerically 
in $O(N)$ time using standard BEM techniques, so here our 
method does not offer a significant computational advantage
over existing techniques.
%%%%%%%%%%%%%%%%%%%%%%%%%%%%%%%%%%%%%%%%%%%%%%%%%%%%%%%%%%%%%%%%%%%%%%
%%%%%%%%%%%%%%%%%%%%%%%%%%%%%%%%%%%%%%%%%%%%%%%%%%%%%%%%%%%%%%%%%%%%%%
%%%%%%%%%%%%%%%%%%%%%%%%%%%%%%%%%%%%%%%%%%%%%%%%%%%%%%%%%%%%%%%%%%%%%%

%%%%%%%%%%%%%%%%%%%%%%%%%%%%%%%%%%%%%%%%%%%%%%%%%%%%%%%%%%%%%%%%%%%%%%
%%%%%%%%%%%%%%%%%%%%%%%%%%%%%%%%%%%%%%%%%%%%%%%%%%%%%%%%%%%%%%%%%%%%%%
%%%%%%%%%%%%%%%%%%%%%%%%%%%%%%%%%%%%%%%%%%%%%%%%%%%%%%%%%%%%%%%%%%%%%%
\section{Force and Torque}
\label{ForceTorqueSection}

\subsection{Force}
The time-average force on $\mathcal{O}$ is obtained by
integrating the Maxwell stress tensor over $\mathcal{S}$:
\begin{align}
 \vb F &= \frac{1}{2}\text{Re}
          \int_{\mathcal{S}} \vb T(\vb x^+) \cdot \vbhat{n}(\vb x) \, dA
\label{Force}
\end{align}
where $\vb x$ lies on $\mathcal{S}$,
$\vbhat{n}$ is the outward-pointing surface normal at
$\vb x$, and the stress tensor $\vb T$ is evaluated
at a point lying just outside the body at $\vb x$,
$$\vb x^+=\lim_{\eta \to 0} \Big(\vb x + \eta\vbhat{n}(\vb x)\Big).$$
The components of $\vb T$ are 
$$
T_{ij} 
=\epsilon E_i^* E_j + \mu H_i^* H_j
  - \frac{\delta_{ij}}{2}\Big[\epsilon |\vb E|^2 + \mu |\vb H|^2 \Big]
$$
where $\epsilon,\mu$ are the material properties of the 
exterior medium in which the object is embedded, 
which we assume to be lossless.

The $i$ component of the integrand of (\ref{Force}) is
(with the convention that we sum over repeated indices)
\numeq{TijNj}
{
 T_{ij} \hat{n}_j 
 =  \epsilon E_i^* \big[ \vb E \cdot \vbhat{n} \big]
    +\mu      H_i^* \big[ \vb H \cdot \vbhat{n} \big]
    -\frac{\hat{n}_i}{2}\Big[\epsilon |\vb E|^2 + \mu |\vb H|^2 \Big]
}
In analogy to equation (\ref{PTotDotN}), we would now like
to rewrite (\ref{TijNj}) in terms of the surface currents
$\vb K$ and $\vb N$; however, whereas Poynting-vector
calculations involve only tangential field components,
equation (\ref{TijNj}) requires knowledge of the 
\textit{normal} field components, which (for points
infinitesimally \textit{outside} the body) are related
to the surface currents according to 
$$ \vb E\cdot\vbhat{n} = \frac{\nabla\cdot\vb K}{i\omega \epsilon},
   \qquad
   \vb H\cdot\vbhat{n} = \frac{\nabla\cdot\vb N}{i\omega \mu}.
$$
Using these relations and equations (\ref{KNDefinition}),
we may write (\ref{TijNj}) in terms of the surface currents: 
\begin{align}
%&=
%  \bigg\{  \big(\vbhat{n} \times \vb N^*\big)_i  
%         + \left(\frac{\nabla\cdot\vb K}{i\epsilon\omega}\right)^* \vbhat{n}_i
%  \bigg\}
%  \bigg\{   \frac{\nabla\cdot\vb K}{i\omega} \bigg\}
%\\
%&\qquad
% +\bigg\{    \big(\vbhat{n} \times \vb K^*\big)_i 
%         + \left(\frac{\nabla\cdot\vb N)}{i\mu\omega}\right)^* \vbhat{n}_i
%  \bigg\}
%  \bigg\{   \frac{\nabla\cdot\vb N}{i\omega}  \bigg\}
%\\
%&\qquad
%  -\frac{\hat{n}_i}{2}
%   \bigg\{  \epsilon\big|\vb N\big|^2 
%           +\frac{\epsilon|\nabla \cdot \vb K|^2}
%                 {|\epsilon|^2\omega^2}
%           +\mu\big|\vb K\big|^2
%           +\frac{\mu|\nabla \cdot \vb N|^2}
%                 {|\mu|^2 \omega^2}
%   \bigg\}
%\\
&\hspace{-0.1in}
 T_{ij}\hat n_j
\label{TijNjFromKN1}\\
= &\frac{1}{i\omega}
   \bigg\{   \big(\vbhat{n} \times \vb N^*\big)_i  (\nabla \cdot \vb K)
           - \big(\vbhat{n} \times \vb K^*\big)_i  (\nabla \cdot \vb N)
   \bigg\}
\nonumber \\
&\quad
  -\frac{\hat{n}_i}{2}
   \left\{   \mu \left[ |\vb K|^2 
           -     \frac{|\nabla \cdot \vb K|^2}{k^2}  \right]
           + \epsilon\left[ |\vb N|^2 
           -         \frac{|\nabla \cdot \vb N|^2}{k^2} \right]
   \right\}.
\nonumber
\intertext{Now insert the expansions (\ref{KNExpansion}):} 
= &\frac{1}{2}\sum_{\alpha\beta} 
    \bigg\{ 
            -\mu \Big[ \hat{n}_i \vb b^*_\alpha \cdot \vb b_\beta 
                     -\frac{1}{k^2} \hat{n}_i \big(\nabla \cdot \vb b_\alpha\big)^*
                                              \big(\nabla \cdot \vb b_\beta\big) 
                  \Big] k_\alpha^* k_\beta 
\nonumber\\
&\hspace{0.45in} 
              -\frac{2}{i\omega} 
                \Big[ \big(\vbhat{n}\times \vb b^*_\alpha\big)_i 
                      \big(\nabla \cdot \vb b_\beta\big)
                \Big] k_\alpha^* n_\beta 
\nonumber\\
&\hspace{0.45in} 
              +\frac{2}{i\omega} 
                \Big[ \big(\vbhat{n}\times \vb b^*_\alpha\big)_i 
                      \big(\nabla \cdot \vb b_\beta\big)
                \Big] n_\alpha^* k_\beta 
\nonumber\\
&\hspace{0.5in} 
      - \epsilon \Big[ \hat{n}_i \vb b^*_\alpha \cdot \vb b_\beta 
                        -\frac{1}{k^2} \hat{n}_i \big(\nabla \cdot \vb b_\alpha\big)^*
                                               \big(\nabla \cdot \vb b_\beta\big) 
                   \Big] n_\alpha^* n_\beta 
    \bigg\}.
\nonumber
%\label{TijNjFromKN2}
%{\tiny
%  -\frac{1}{2}
%  \sum_{\alpha\beta} 
%  %--------------------------------------------------------------------%
%  \left( 
%     \!\!\!
%     \begin{array}{c} 
%     k_\alpha 
%     \\[5pt] 
%     n_\alpha 
%     \end{array}
%     \!\!\!
%  \right)^\dagger
%  %--------------------------------------------------------------------%'
%  \left(\begin{array}{cc} 
%    \frac{\epsilon^* \vbhat{n}_i}{4\omega^2} 
%    \big(\nabla \cdot \vb b_\alpha^*\big)
%    \big(\nabla \cdot \vb b_\beta\big)
%    -\mu \vbhat{n}_i \vb b_\alpha^* \cdot \vb b_\beta
%   &
%   \frac{1}{i\omega} 
%   (\vbhat{n}\times \vb b_\alpha^*)(\nabla \cdot \vb b_\beta)
%   \\[5pt]
%   \frac{1}{i\omega} 
%   (\vbhat{n}\times \vb b_\alpha^*)(\nabla \cdot \vb b_\beta)
%   &
%    \frac{\mu^* \vbhat{n}_i }{4\omega^2} 
%    \big(\nabla \cdot \vb b_\alpha^*\big)
%    \big(\nabla \cdot \vb b_\beta\big)
%    -\epsilon \vbhat{n}_i \vb b_\alpha^* \cdot \vb b_\beta
%  \end{array}\right)
%  %--------------------------------------------------------------------%
%  \left(
%     \!\!\!
%     \begin{array}{c} 
%     k_\alpha 
%     \\[5pt] 
%     n_\alpha 
%     \end{array}
%     \!\!\!
%  \right)
%}
\end{align}
Finally, inserting into (\ref{Force}) and evaluating the surface
integrals over basis functions yields a bilinear product expression,
analogous to (\ref{PAbsFormula}), for the $i$-directed
force on the object:
\numeq{Fi}
{\boxed{F_i
= \frac{1}{2}\text{ Re } \int T_{ij}\hat n_j dA 
= \frac{1}{4}\text{ Re } \vb c^\dagger \vb O\IF \vb c}
}
where the entries of the matrix $\vb O\IF$ describe
various types of overlap integrals between the basis
functions: 
%====================================================================%
\begin{align}
  &O\IFAB
\label{ForceBlockMatrix} \\ 
&\,\,
=\left(\begin{array}{cc}
       -\mu
             \Big[  O\IFAB^{\bullet} 
                   -\frac{1}{k^2}O\IFAB^{\nabla\nabla} 
             \Big]
 &
 -\frac{2}{i\omega} O\IFAB^{\times\nabla}
\\[5pt]
  \frac{2}{i\omega} O\IFAB^{\times\nabla}
 &
    -\epsilon
              \Big[  O\IFAB^{\bullet} 
                   -\frac{1}{k^2}O\IFAB^{\nabla\nabla} 
              \Big]
 \end{array}\right)
\nonumber
\end{align}
%====================================================================%
where the overlap integrals are 
\begin{align*}
 O^{\bullet}\IFAB
&\equiv \int 
   \vbhat{n}_i(\vb x) 
   \vb b_\alpha^*(\vb x) \cdot \vb b_\beta(\vb x) 
   d\vb x
\\
 O^{\nabla \nabla}\IFAB
&\equiv \int 
   \vbhat{n}_i(\vb x)
   \big[\nabla \cdot \vb b_\alpha^*(\vb x)\big] 
   \big[\nabla \cdot \vb b_\beta(\vb x)\big]
   d\vb x
\\
 O^{\times \nabla}\IFAB
&\equiv \int 
   \big[\vbhat{n}(\vb x)\times \vb b_\alpha^*(\vb x)\big]_i
   \big[\nabla \cdot \vb b_\beta(\vb x)\big]
   d\vb x.
\end{align*}
%=====================================================================
As was true for the power formulas 
(\ref{PAbsFormula}), (\ref{PScatFormula}), and (\ref{PTotFormula}),
if we work in a basis of $N$ localized functions then the matrix 
(\ref{ForceBlockMatrix}) is sparse and the cost of evaluating 
(\ref{Fi}) scales like $O(N).$

%%%%%%%%%%%%%%%%%%%%%%%%%%%%%%%%%%%%%%%%%%%%%%%%%%%%%%%%%%%%%%%%%%%%%%
%%%%%%%%%%%%%%%%%%%%%%%%%%%%%%%%%%%%%%%%%%%%%%%%%%%%%%%%%%%%%%%%%%%%%%
%%%%%%%%%%%%%%%%%%%%%%%%%%%%%%%%%%%%%%%%%%%%%%%%%%%%%%%%%%%%%%%%%%%%%%
\subsection{Torque}

The time-average torque on $\mathcal{O}$ is 
%====================================================================%
\begin{align}
 \boldsymbol{\mathcal{T}} 
 &= \frac{1}{2}\text{Re}
    \int_{\mathcal{S}} \Big[ (\vb x - \vb x_0) \times \vb T(\vb x^+)\Big]
                        \vbhat{n}(\vb x) \, dA
\nonumber \\
\intertext{or, in componentwise notation,}
 \mathcal{T}_i
 &= \frac{1}{2}\text{Re}
    \int_{\mathcal{S}} \varepsilon_{ijk} (\vb x-\vb x_0)_j
                        T_{k\ell}(\vb x^+) n_{\ell}(\vb x) \, dA
\label{Torque}
\end{align}
%====================================================================%
where $\vb x_0$ is the origin about which we figure the torque
and $\varepsilon$ is the Levi-Civita symbol.
A calculation analogous to the above yields an 
expression analogous to (\ref{Fi}) for the torque:
\numeq{Ti}
{ \boxed{\mathcal{T}_i
= \frac{1}{4}\text{ Re } \vb c^\dagger \vb O\IT \vb c}
}
where elements of the torque overlap matrix $\vb O\IT$ 
have the structure
%====================================================================%
\begin{align}
 &O\ITAB
\label{TorqueBlockMatrix} \\
 &\,\,
=\left(\begin{array}{cc}
       -\mu \Big[  O\ITAB^{\bullet} 
                    -\frac{1}{k^2}O\ITAB^{\nabla\nabla} 
              \Big]
 &
 -\frac{2}{i\omega} O\ITAB^{\times\nabla}
\\[5pt]
  \frac{2}{i\omega} O\ITAB^{\times\nabla}
 &
 -\epsilon
              \Big[  O\ITAB^{\bullet} 
                   -\frac{1}{k^2}O\ITAB^{\nabla\nabla} 
              \Big]
 \end{array}\right)
\nonumber
\end{align}
%====================================================================%
with a modified set of overlap integrals:
%====================================================================%
%====================================================================%
%====================================================================%
\begin{align*}
 O^{\bullet}\ITAB
&\equiv \int 
   \Big[ \big(\vb x - \vb x_0\big) \times \vbhat{n}(\vb x) \Big]_i
   \vb b_\alpha^*(\vb x) \cdot \vb b_\beta(\vb x) 
   d\vb x
\\
 O^{\nabla \nabla}\ITAB
&\equiv \int 
   \Big[ \big(\vb x - \vb x_0\big) \times \vbhat{n}(\vb x) \Big]_i
   \big[\nabla \cdot \vb b_\alpha^*(\vb x)\big] 
   \big[\nabla \cdot \vb b_\beta(\vb x)\big]
   d\vb x
\\
 O^{\times \nabla}\ITAB
&\equiv \int 
   \Big\{ \big(\vb x - \vb x_0\big) \times 
          \Big[ \vbhat{n}(\vb x)\times \vb b_\alpha^*(\vb x)\Big]
   \Big\}_i
   \big[\nabla \cdot \vb b_\beta(\vb x)\big]
   d\vb x.
\end{align*}
Again,
if we work in a basis of $N$ localized functions then the matrix 
(\ref{TorqueBlockMatrix}) is sparse and the cost of evaluating 
(\ref{Ti}) scales like $O(N).$
%%%%%%%%%%%%%%%%%%%%%%%%%%%%%%%%%%%%%%%%%%%%%%%%%%%%%%%%%%%%%%%%%%%%%%
%%%%%%%%%%%%%%%%%%%%%%%%%%%%%%%%%%%%%%%%%%%%%%%%%%%%%%%%%%%%%%%%%%%%%%
%%%%%%%%%%%%%%%%%%%%%%%%%%%%%%%%%%%%%%%%%%%%%%%%%%%%%%%%%%%%%%%%%%%%%%

%%%%%%%%%%%%%%%%%%%%%%%%%%%%%%%%%%%%%%%%%%%%%%%%%%%%%%%%%%%%%%%%%%%%%%
%%%%%%%%%%%%%%%%%%%%%%%%%%%%%%%%%%%%%%%%%%%%%%%%%%%%%%%%%%%%%%%%%%%%%%
%%%%%%%%%%%%%%%%%%%%%%%%%%%%%%%%%%%%%%%%%%%%%%%%%%%%%%%%%%%%%%%%%%%%%%
\section{Computational Examples}
\label{ExamplesSection}

\subsection{Energy and Momentum Transfer in Mie Scattering}
%####################################################################%
\begin{figure}
\begin{center}
\resizebox{0.5\textwidth}{!}{\includegraphics{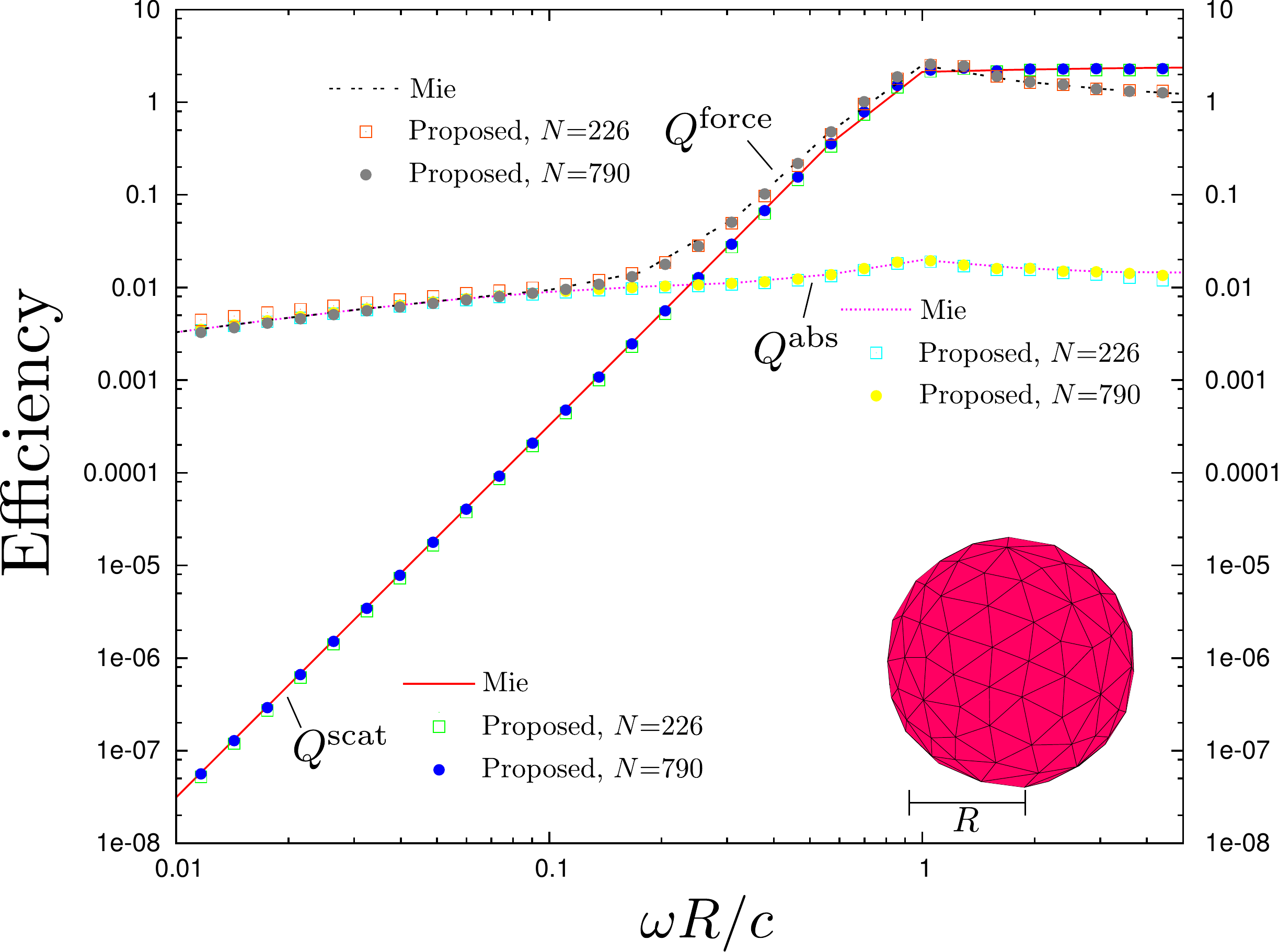}}
\caption{Absorption, scattering, and force efficiencies for a spherical
         gold nanoparticle of radius $R=1\,\mu$m irradiated by a 
         linearly-polarized plane wave. (Efficiencies are cross sections
         normalized by the geometric cross section, $Q=\frac{\sigma}{\pi R^2}.$
         The absorption and scattering cross sections are the total absorbed
         and scattered power normalized by the incident power flux.
         The force cross section is the force normalized by the
         incident momentum flux.) Solid and dashed lines 
         are the results of Mie theory. Squares (circles) are the
         results of calculations using the method proposed in this
         paper for a sphere discretized into $N=226$ ($N=790$)
         flat triangular panels with surface currents described
         by RWG basis functions. The inset shows the $N=226$ surface
         mesh.}
\label{SphereDataFigure}
\end{center}
\end{figure}
%####################################################################%

We first validate our new formulas by using them to reproduce
well-known results for a dielectric sphere irradiated by a 
plane wave (Mie scattering). Analytical formulas for the
absorbed power, scattered power, and force (radiation pressure)
in this case may be found in standard 
textbooks~\cite{BohrenHuffman1983}. 

Figure \ref{SphereDataFigure} plots
the efficiencies of scattering, absorption, and $z$-directed
force on a spherical nanoparticle of radius $R=1\,\mu$m,
irradiated by a linearly-polarized $z$-traveling plane wave 
with electric field 
$\vb E\sups{inc}(\vb x, \vb t)=E_0\vbhat{x} e^{i(kz-\omega t)}.$
The nanoparticle is composed of gold, modeled as a lossy
dielectric with frequency-dependent permittivity 
\numeq{EpsilonGold}
{ \epsilon\subs{gold}
   =
   \epsilon_0\left( 1 - \frac{\omega_p^2}{\omega(\omega+i\gamma)}\right)
}
with $\{\omega_p,\gamma\}=\{1.37\cdot 10^{16}, 5.32\cdot 10^{13}\}$
rad / sec.
Solid lines indicate the results of Mie theory~\cite{BohrenHuffman1983},
while squares (circles) indicate data points obtained using the 
method proposed in this paper using RWG basis functions
for a sphere discretized into $N$=226 ($N$=790) trianglular 
panels. (The figure inset shows the $N$=226 mesh.) The absorption 
and scattering 
cross sections $\sigma\sups{abs,scat}$ are obtained by 
dividing the total absorbed and scattered power by the 
incident power flux, 
$\sigma\sups{abs,scat}=P\sups{abs,scat}/P\sups{inc}$,
where $P\sups{inc}=\frac{|E_0|^2}{2Z_0}$ for a plane wave
in vacuum and $P\sups{abs,scat}$ are computed using 
equations (\ref{PAbsFormula}) and (\ref{PScatFormula}).
%[or (\ref{AlternativePScatFormula}]. 
The force
cross section is obtained by dividing the total 
$z$-directed force on the particle by the incident 
momentum flux, $\sigma\sups{force}=F_z/F\sups{inc}$,
where $F_z$ is computed using equation (\ref{Fi}) 
and $F\sups{inc}=\frac{|E_0|^2}{2Z_0 c}$ for a 
plane wave in vacuum.
Efficiencies $Q\sups{abs,scat,force}$ are obtained
from cross sections $\sigma\sups{abs,scat,force}$ by dividing
by the geometrical cross section presented by the 
particle, $Q=\sigma/(\pi R^2).$

On the scale of this graph, errors due to finite 
mesh sizes are discernible only at the highest and
lowest frequencies. The former are just the usual
finite-mesh inaccuracies that arise in all BEM 
schemes when the wavelength of light is comparable 
to or smaller than the panel sizes. On the other hand, 
finite-mesh errors at \textit{low} frequencies in 
quantities such as the force arise because 
the stress-tensor integral over the body surface 
implicit in the VMVP formula (\ref{Fi}) exhibits,
at low frequencies, large cancellations from different 
regions of the body surface that require fine meshing 
to resolve accurately.

There is of course no torque on a homogeneous spherical
body irradiated by a plane wave. However, a nonzero torque
may develop when the incident field is a finite-width
laser beam whose center is displaced from the sphere
center, and this situation may be studied numerically 
using a generalized Mie technique that allows arbitrary
incident fields~\cite{libSphericalScattering}, with the
torque computed by integrating the Maxwell stress tensor
over a bounding surface. We have confirmed that the results
of this procedure agree with results predicted by our
formula (\ref{Ti}) for the torque on a homogeneous
sphere irradiated by an off-center laser beam.

%%%%%%%%%%%%%%%%%%%%%%%%%%%%%%%%%%%%%%%%%%%%%%%%%%%%%%%%%%%%%%%%%%%%%%
%%%%%%%%%%%%%%%%%%%%%%%%%%%%%%%%%%%%%%%%%%%%%%%%%%%%%%%%%%%%%%%%%%%%%%
%%%%%%%%%%%%%%%%%%%%%%%%%%%%%%%%%%%%%%%%%%%%%%%%%%%%%%%%%%%%%%%%%%%%%%
\subsection{Force and Torque on a Chiral Nanoparticle in a 
            Circularly-Polarized Field}
%####################################################################%
\begin{figure}
\begin{center}
\resizebox{0.5\textwidth}{!}{\includegraphics{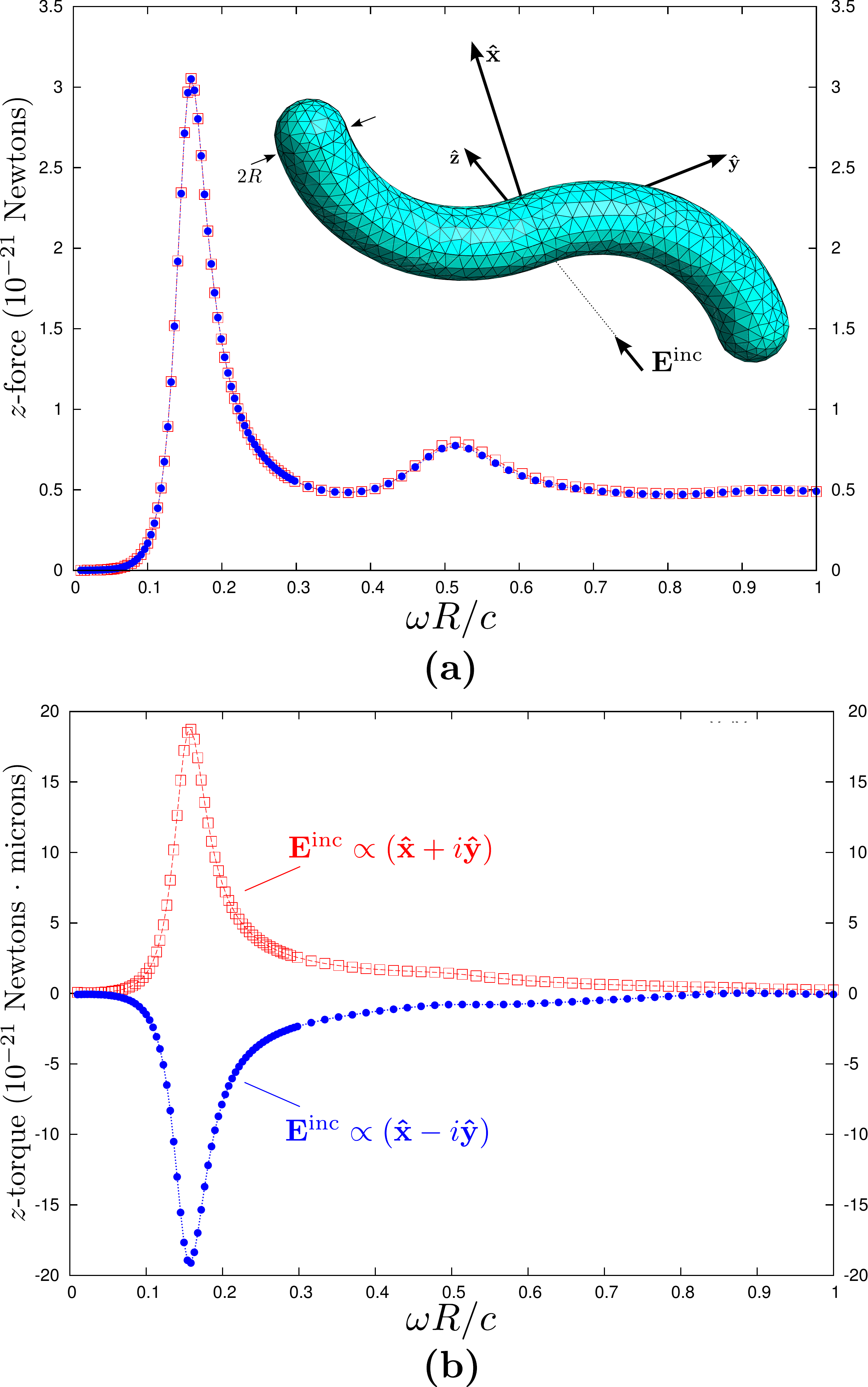}}
\caption{$z$-directed force \textbf{(a)} and torque \textbf{(b)} on a 
         chiral gold nanoparticle
         illuminated from below by a plane wave traveling in the positive $z$ 
         direction with left circular polarization (red data points) or
         right circular polarization (blue data points).
         The \textit{force} is the same for the two incident polarizations, while
         the \textit{torque} changes sign when the incident polarization is reversed.}
\label{ChiralParticleFigure}
\end{center}
\end{figure}
%####################################################################%

We next consider a chiral nanoparticle illuminated by 
a circularly-polarized plane wave. The particle is
depicted in the inset of Figure \ref{ChiralParticleFigure}\textbf{(a)};
it is constructed from two quarter-length sections of a
torus of \{inner, outer\} radii $\{R,5R\}$ (here $R=1\mu$m)
with one section rotated through $\pi$ degrees
and both ends capped with hemispherical endcaps.
The particle is composed of gold with permittivity 
given by equation (\ref{EpsilonGold}). The particle is 
illuminated from below by a plane wave traveling in the 
positive $\vbhat{z}$ direction and either left- or 
right-circular polarization, i.e. we have 
%====================================================================%
$$\vb E\sups{inc}(\vb x, t)=E_0(\vbhat{x} \pm i\vbhat{y})e^{i(kz-\omega t)}$$
%====================================================================%
with $k=\omega/c$ and $E_0=1$ V/m. 
Figures \ref{ChiralParticleFigure}\textbf{(a)} and 
\ref{ChiralParticleFigure}\textbf{(b)} respectively plot the force and 
torque on the particle as a function of frequency for both 
right-circularly polarized incident fields (blue circles) and
left-circularly polarized incident fields (red squares).
The \textit{force} on the particle is the same for the two incident 
polarizations, while the \textit{torque} changes sign
when the incident polarization is reversed. 

%%%%%%%%%%%%%%%%%%%%%%%%%%%%%%%%%%%%%%%%%%%%%%%%%%%%%%%%%%%%%%%%%%%%%%
%%%%%%%%%%%%%%%%%%%%%%%%%%%%%%%%%%%%%%%%%%%%%%%%%%%%%%%%%%%%%%%%%%%%%%
%%%%%%%%%%%%%%%%%%%%%%%%%%%%%%%%%%%%%%%%%%%%%%%%%%%%%%%%%%%%%%%%%%%%%%
\subsection{Frequency-Dependent Direction of Rotation on a Pair 
            of Chiral Nanoparticles in a Linearly-Polarized Field}
%####################################################################%
\begin{figure}
\begin{center}
\resizebox{0.5\textwidth}{!}{\includegraphics{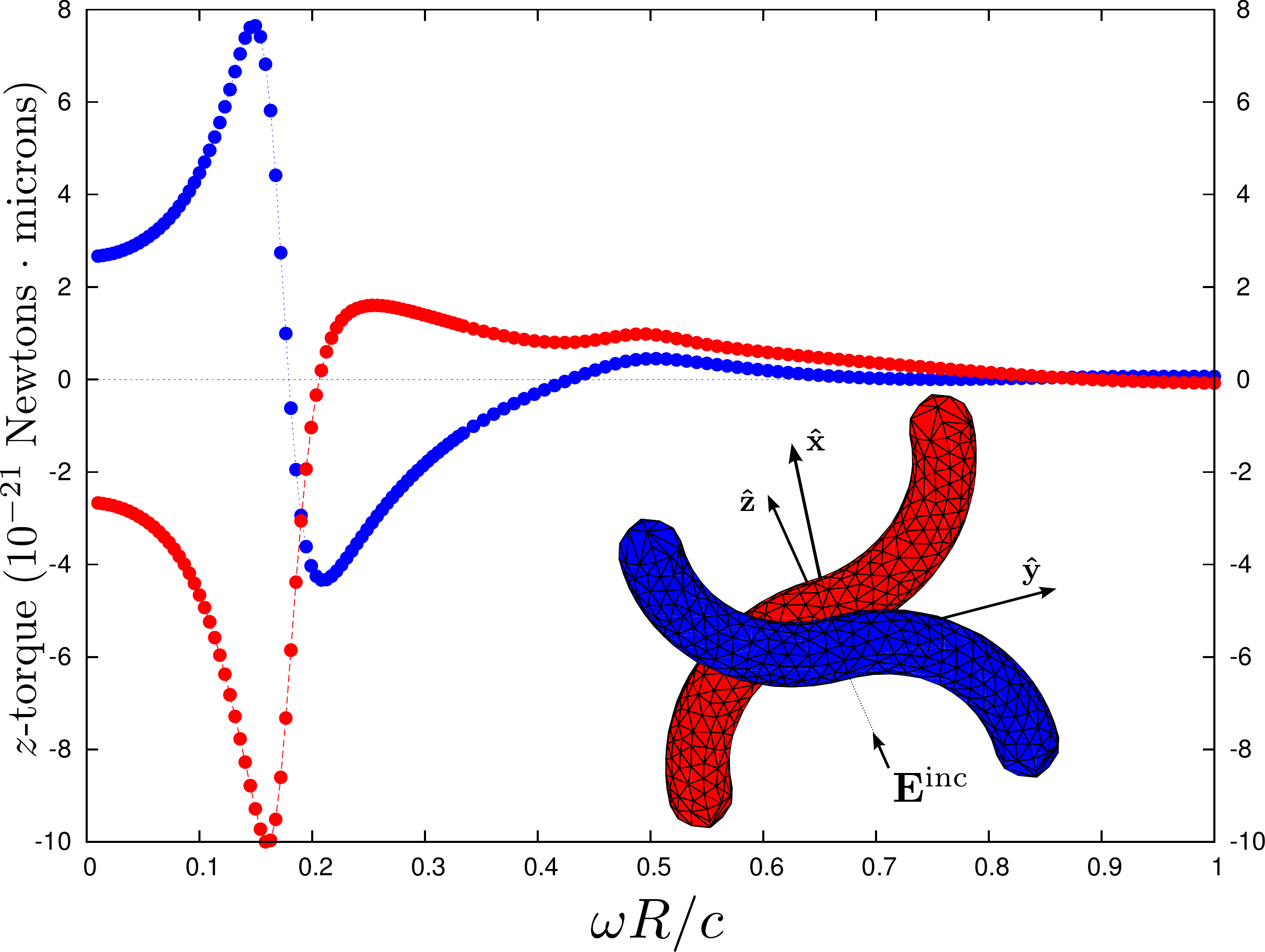}}
\caption{$z$-directed torque on the lower (blue data points) and upper 
         (red data points) members of a pair of chiral gold nanoparticles 
         with opposite chirality separated by a surface--surface separation
         equal to the inner radius of the particle. The particle pair 
         is illuminated from below by a linearly-polarized plane wave 
         traveling in the positive $z$ direction. Now not only the 
         magnitudes but also the \textit{signs} of the torques on both 
         particles are frequency dependent, with counter-rotating and 
         co-rotating frequency regimes.}
\label{ChiralParticlePairFigure}
\end{center}
\end{figure}
%####################################################################%

In the previous two examples---which involved isolated nanoparticles---the
PFTs could in principle have been computed by the standard 
technique of integrating the PV or MST over a distant bounding surface
surrounding the object, in which case the computation would
involve only far fields and could have been performed without
great difficulty using existing methods.

The method we propose in this paper really comes into its own for
geometries involving two or more closely separated bodies for which 
we wish separately to resolve the PFTs. In such cases, any bounding 
surface surrounding only one of the bodies in question must unavoidably 
lie close to both bodies, and the stress-tensor integral over 
that surface will be particularly cumbersome, whereas the 
technique proposed in this paper is straightforward.
An example of such a 
geometry is provided by pairing the chiral gold nanoparticle
of the previous section with a partner gold nanoparticle of
opposite chirality, separated by a surface-surface
separation $d=1\,\mu$m (equal to the inner particle radius),
and irradating the pair from
below with a \textit{linearly}, not circularly, 
polarized plane wave,
$\vb E\sups{inc}(\vb x, \vb t)=E_0\vbhat{x} e^{i(kz-\omega t)}.$
Figure \ref{ChiralParticlePairFigure} plots the torque
on each particle, with the blue (red) data points
corresponding to the lower (upper) particle.
Now not only the magnitudes but also the \textit{signs}
of the torques on the two particles are frequency-dependent,
with the particles either co-rotating, counter-rotating, or 
not rotating depending on the frequency.
%%%%%%%%%%%%%%%%%%%%%%%%%%%%%%%%%%%%%%%%%%%%%%%%%%%%%%%%%%%%%%%%%%%%%%
%%%%%%%%%%%%%%%%%%%%%%%%%%%%%%%%%%%%%%%%%%%%%%%%%%%%%%%%%%%%%%%%%%%%%%
%%%%%%%%%%%%%%%%%%%%%%%%%%%%%%%%%%%%%%%%%%%%%%%%%%%%%%%%%%%%%%%%%%%%%%

%%%%%%%%%%%%%%%%%%%%%%%%%%%%%%%%%%%%%%%%%%%%%%%%%%%%%%%%%%%%%%%%%%%%%%
%%%%%%%%%%%%%%%%%%%%%%%%%%%%%%%%%%%%%%%%%%%%%%%%%%%%%%%%%%%%%%%%%%%%%%
%%%%%%%%%%%%%%%%%%%%%%%%%%%%%%%%%%%%%%%%%%%%%%%%%%%%%%%%%%%%%%%%%%%%%%
\section{Summary and Outlook}
\label{ConclusionsSection}

In this paper we derived concise and computationally efficient 
new formulas for the power, force, and torque on 
material bodies in classical scattering problems.
These formulas may be viewed as the classical analogues of
the ``fluctuating-surface-current'' formulas for 
Casimir forces and other fluctuation 
phenomena~\cite{Reid2012A, Reid2012B, Rodriguez2013},
and they may be derived and understood on the basis of
at least two complementary physical pictures.
We validated our new formulas by confirming that they
reproduce analytical results for spheres, then used them
to compute optical forces and torques in complex geometries
that would be inconvenient to address using other methods.

In Section \ref{PowerSection} we noted that our formula
(\ref{PTotFormula2}) for the total power had the advantage
of manifest non-negativity, which follows from the 
negative semidefiniteness of the BEM matrix $\vb M$~\cite{Rodriguez2013}.
A similar statement holds for the alternative [$O(N^2)$]
scattered power formula, equation (\ref{AlternativePScatFormula}).
However, at present we are unaware of any demonstration
that our our $O(N)$ formulas (\ref{PAbsFormula})
and (\ref{PScatFormula}) share this manifest
non-negativity property. It would be of interest 
to prove that these expressions for non-negative
physical quantities are in fact provably non-negative---or
to derive alternative $O(N)$ expressions that are.

In closing, we emphasize that all formulas derived in this 
paper are implemented in {\sc scuff-em}, a free, open-source 
software implementation of the BEM in the EFIE and PMCHW
formulations~\cite{scuff-em}.

\section*{Acknowledgments}

This work was supported in part by the Defense Advanced Research
Projects Agency (DARPA) under grant N66001-09-1-2070-DOD, by the Army
Research Office through the Institute for Soldier Nanotechnologies
(ISN) under grant W911NF-07-D-0004, and by the AFOSR Multidisciplinary
Research Program of the University Research Initiative (MURI) for
Complex and Robust On-chip Nanophotonics under grant FA9550-09-1-0704.
%%%%%%%%%%%%%%%%%%%%%%%%%%%%%%%%%%%%%%%%%%%%%%%%%%%%%%%%%%%%%%%%%%%%%%
%%%%%%%%%%%%%%%%%%%%%%%%%%%%%%%%%%%%%%%%%%%%%%%%%%%%%%%%%%%%%%%%%%%%%%
%%%%%%%%%%%%%%%%%%%%%%%%%%%%%%%%%%%%%%%%%%%%%%%%%%%%%%%%%%%%%%%%%%%%%%

%%%%%%%%%%%%%%%%%%%%%%%%%%%%%%%%%%%%%%%%%%%%%%%%%%%%%%%%%%%%%%%%%%%%%%
%%%%%%%%%%%%%%%%%%%%%%%%%%%%%%%%%%%%%%%%%%%%%%%%%%%%%%%%%%%%%%%%%%%%%%
%%%%%%%%%%%%%%%%%%%%%%%%%%%%%%%%%%%%%%%%%%%%%%%%%%%%%%%%%%%%%%%%%%%%%%
\appendices

%%%%%%%%%%%%%%%%%%%%%%%%%%%%%%%%%%%%%%%%%%%%%%%%%%%%%%%%%%%%%%%%%%%%%%
%%%%%%%%%%%%%%%%%%%%%%%%%%%%%%%%%%%%%%%%%%%%%%%%%%%%%%%%%%%%%%%%%%%%%%
%%%%%%%%%%%%%%%%%%%%%%%%%%%%%%%%%%%%%%%%%%%%%%%%%%%%%%%%%%%%%%%%%%%%%%
\section{Equivalence-principle interpretation of power formulas}
\label{AlternativeScatteredPowerAppendix} 

Equations (\ref{PAbsFormula}) and (\ref{PTotFormula}) for the
absorbed and total power were derived in Section \ref{PowerSection}
by integrating the Poynting vector over a bounding surface to 
compute energy flow into and out of a spatial volume. It is
interesting to note that precisely the \textit{same} formulas
may be derived on the basis of an alternative picture in which
the principle of equivalence~\cite{Harrington1961,Chen1989} is 
invoked to eliminate the material body, with questions of energy 
transfer recast as questions of \textit{work} done on currents by 
fields or vice versa (work done on fields by currents).
In addition to providing an appealing physical picture
that complements the viewpoint of Section \ref{PowerSection},
this approach has the advantage of furnishing a new VMVP formula 
for the \textit{scattered} power which, though less computationally 
efficient than equation (\ref{PScatFormula}), avoids the numerical 
inaccuracies that befall that formula in cases where
the extinction is dominated by absorption.

[Similar procedures for computing absorbed or scattered powers by
evaluating the work done on currents by fields---or vice versa---were
recently employed in \citeasnoun{Hashemi2012} in the context of
cloaking bandwidth studies. A distinction is that, whereas
there the currents were \textit{physical} volume currents
on which the work done has immediate physical relevance,
here the currents are \textit{effective}---that is,
fictitious---surface currents. This renders the corresponding
physical interpretation somewhat more subtle, although
the mathematical equivalence between the calculated
work and the physical power is no less exact here than
there.]

For simplicity we consider the case of a single compact 
homogeneous material body embedded in a homogeneous exterior
medium and irradiated by an incident field whose sources 
lie in the exterior medium. A standard 
approach to formulating surface-integral equations is to
solve for surface currents by demanding that the scattered
fields they radiate satisfy the boundary conditions
at all media interfaces~\cite{Medgyesi1994}. 
The equivalence-principle approach
(see, for instance, Figure 3-9[b] of \citeasnoun{Harrington1961})
traverses an alternative logical sequence to arrive at
the same surface-integral equations. In this picture, 
\textbf{(a)} the material object is absent, and hence all 
space is permeated simply by vacuum; \textbf{(b)} a distribution 
of electric and magnetic currents, created and sustained by 
unspecified means, exists throughout the two-dimensional region
of space that would be occupied by the object surface if it were
present (we will refer to this region as the ``object surface''
even though no object is present) and flows tangentially to 
that surface.  \textbf{(c)} The total electromagnetic field 
everywhere in space is the sum of the incident field plus the 
field radiated by these currents (the ``scattered'' field).
The scattered field at any point---including points in the
interior of the region that would be occupied by the object
if it were present (the ``object volume'')---is computed by
convolving the surface-current distribution with the 
dyadic Green's functions (DGFs) for the \textit{exterior}
medium.
\textbf{(d)} The total fields computed in this
way turn out to yield 
\textbf{(i)} exactly zero for points inside the object volume,
\textbf{(ii)} exactly the correct total fields for points 
outside the object volume,
\textbf{(iii)} exactly \textit{one-half} the correct total 
fields for points on the object surface~\cite{Chen1989}.

In this picture, there is no absorbing medium 
anywhere in space, so there can be no volume 
power absorption. Instead, the absorbed, 
total, and scattered power are computed by 
considering the work done on the surface currents
by the fields, to wit:

\begin{itemize}
 \item the \textit{absorbed} power is equal to the 
       work done on the surface currents by the 
       \textit{total} fields:
%====================================================================%
       \numeq{PAbsAlternative}
       {P\sups{abs}=\frac{1}{2}\text{Re } \oint_{\partial \mc O}
                    \Big\{  \vb K^*(\vb x) \cdot \vb E\sups{tot}(\vb x)
                           +\vb N^*(\vb x) \cdot \vb H\sups{tot}(\vb x)
                    \Big\}d\vb x
       }
%====================================================================%
       
 \item the \textit{total} power (extinction) is equal 
       to the work done on the surface currents by the 
       \textit{incident} fields:
%====================================================================%
       \numeq{PTotAlternative}
       {P\sups{tot}=\frac{1}{2}\text{Re } \oint_{\partial \mc O}
                    \Big\{  \vb K^*(\vb x) \cdot \vb E\sups{inc}(\vb x)
                           +\vb N^*(\vb x) \cdot \vb H\sups{inc}(\vb x)
                    \Big\}d\vb x
       }
%====================================================================%
 \item the \textit{scattered} power is equal to the 
       work done \textit{by} the surface currents (or by
       the unspecified mechanism which sustains them)
       on the \textit{scattered} fields:
%====================================================================%
       \begin{align}
        P\sups{scat}&=-\frac{1}{2}\text{Re } \oint_{\partial \mc O}
                       \Big\{  \vb K^*(\vb x) \cdot \vb E\sups{scat}(\vb x)
       \nn
        &\hspace{1in}  +\vb N^*(\vb x) \cdot \vb H\sups{scat}(\vb x)
                       \Big\}d\vb x.
       \label{PScatAlternative}
       \end{align}
%====================================================================%
       (The minus sign arises because we are considering
       work done by currents on fields, not the other
       way around as usual).
\end{itemize}

Inserting equations (\ref{KNExpansion}) and (\ref{RHSDefinition}) into
equations (\ref{PAbsAlternative}) and (\ref{PTotAlternative})
and evaluating the integrals over basis functions immediately
recovers equations (\ref{PAbsFormula}) and (\ref{PTotFormula}).

On the other hand, the VMVP formula that arises from 
(\ref{PScatAlternative}) differs from the VMVP formula 
(\ref{PScatFormula}) for the scattered power. To see this, 
note that the scattered fields on the object surface are 
obtained by convolving the surface currents with the DGFs 
for the exterior medium:
%====================================================================%
\numeq{EScatEquation}
{
 \left(\begin{array}{c}
 \vb E\sups{scat} \\ 
 \vb H\sups{scat}
 \end{array}\right)
=
 +
 \frac{1}{2}
 \left(\begin{array}{cc}
 \vbGamma\subs{ext}\supt{EE} & \vbGamma\subs{ext}\supt{EM} \\ 
 \vbGamma\subs{ext}\supt{ME} & \vbGamma\subs{ext}\supt{MM}
 \end{array}\right)
  \star
 \left(\begin{array}{c}
 \vb K \\ 
 \vb N
 \end{array} \right)
}
%====================================================================%
Here e.g. $\vbGamma\subs{ext}\supt{EM}(\vb x, \vb x^\prime)$ represents
the electric field at a point $\vb x$ due to a unit-strength pointlike 
source of magnetic current at $\vb x^\prime$ (the ``ext'' designation
indicates that these are the DGFs of the exterior medium), $\star$ denotes 
convolution, and the factor of $\frac{1}{2}$ arises because we are 
evaluating the fields on the body surface~\cite{Chen1989}.
Inserting (\ref{EScatEquation}) into (\ref{PScatAlternative}), 
we have 
$$
 P\sups{scat} =
 -\frac{1}{4}\text{Re } \oint_{\partial \mc O}
        \left(\begin{array}{c}
        \vb K \\ \vb N
        \end{array}\right)^\dagger
%        \left(
%               \vphantom{ \left(\begin{array}{c} \vb K \\ \vb N \end{array}\right) }
        \text{\large{$\vbGamma\subs{ext}$}}
%        \right)
        \star
        \left(\begin{array}{c}
        \vb K \\ \vb N
        \end{array}\right)
        d\vb x
$$
where $\vbGamma\subs{ext}$ is the 6$\times$6 DGF that
enters equation (\ref{EScatEquation}).
Finally, inserting the surface-current expansions 
(\ref{KNExpansion}) and evaluating the integrals 
over basis functions
yields a VMVP expression for the scattered power:
\numeq{AlternativePScatFormula}
{
 \boxed{
 P\sups{scat} = 
 -\frac{1}{4}\text{Re } \vb c^\dagger \vb M\subs{ext} \vb c
       }
}
where, as in Section \ref{PowerSection}, 
$\vb c$ is the vector of surface-current expansion coefficients
and the elements
of the matrix $\vb M\subs{ext}$ are the interactions of the basis 
functions mediated by $\vbGamma\subs{ext}$:
\numeq{MExt}
{
   M_{\text{\scriptsize{ext}},\alpha\beta}
   =
   \left(\begin{array}{cc} 
   M_{\text{\scriptsize{ext}},\alpha\beta}\supt{EE} &
   M_{\text{\scriptsize{ext}},\alpha\beta}\supt{EM} \\
   M_{\text{\scriptsize{ext}},\alpha\beta}\supt{ME} &
   M_{\text{\scriptsize{ext}},\alpha\beta}\supt{MM} 
  \end{array}\right)
}
with
\numeq{MExtElement}
{
   M_{\text{\scriptsize{ext}},\alpha\beta}^{pq}
   =
   \Big\langle \vb b_\alpha , \vbGamma\subs{ext}^{pq} \star \vb b_\beta
   \Big\rangle
}
for $p,q\in$\{E,M\}. Like equation (\ref{PTotFormula2}),
equation (\ref{AlternativePScatFormula}) has the advantage
of being a manifestly nonnegative expression for the
physically nonnegative 
scattered power; again this follows from the negative 
semidefiniteness of the matrix $\vb M\subs{ext}$~\cite{Rodriguez2013}.

Computing the elements (\ref{MExtElement}) entails no additional
work beyond what is necessary to assemble the BEM system;
indeed, the matrix $\vb M$ in equation (\ref{BEMSystem}) is (in the 
PMCHW formulation) just the sum of $\vb M\subs{ext}$ 
plus its counterpart matrix involving the interactions of the basis
functions mediated by the DGFs of the medium \textit{interior}
to the scattering object,
%====================================================================%
\numeq{MTotMExtMInt}{\vb M = \vb M\subs{ext} + \vb M\subs{int}}
%====================================================================%
where $\vb M\subs{int}$ is computed as in 
(\ref{MExt}-\ref{MExtElement}) with the replacement 
$\vbGamma\subs{ext}\to\vbGamma\subs{int}$. Hence
the work of computing the matrix in 
(\ref{AlternativePScatFormula}).
is already done as soon as we have assembled the BEM 
system (\ref{BEMSystem}).

The VMVP formula for the scattered power derived
in Section \ref{PowerSection} [equation (\ref{PScatFormula})]
obtains the scattered power as the difference between
the total power (extinction) and the absorbed power.
In cases where the total power is dominated by
absorption---such as for the gold sphere of
Figure \ref{SphereDataFigure} at low frequencies---this
amounts to computing a small number as the difference
between two nearby large numbers and hence invites
numerical inaccuracy.
Equation (\ref{AlternativePScatFormula}) provides an alternative 
VMVP formula for the scattered power that avoids this 
difficulty. The drawback is that the matrix in (\ref{MExtElement}) 
is dense, so the cost of evaluating (\ref{AlternativePScatFormula}) 
scales like $O(N^2)$ instead of the $O(N)$ scaling of (\ref{PScatFormula}).

By combining equations
(\ref{PTotFormula2}), (\ref{AlternativePScatFormula}), and
(\ref{MTotMExtMInt}), we obtain an alternative VMVP formula 
for the \textit{absorbed} power: 
%====================================================================%
\begin{align}
  P\sups{abs}
&= P\sups{tot} - P\sups{scat} \nn
&= -\frac{1}{4} \vb c^\dagger \vb v
   -\frac{1}{4}\text{Re }\vb c^\dagger 
    \Big[ \vb M - \vb M\subs{ext} \Big] \vb c \nn
&= -\frac{1}{4}\text{Re }\vb c^\dagger\Big[\vb v + \vb M\subs{int} \vb c\Big].
\label{AlternativePAbsFormula}
\end{align}
%====================================================================%
Equation (\ref{AlternativePAbsFormula}) complements
(\ref{PAbsFormula}) in the same way that 
Equation (\ref{AlternativePScatFormula}) complements (\ref{PScatFormula}):
It is an $O(N^2)$ formula for which the matrix elements are
already computed by the usual BEM assembly process and hence 
require no additional work to calculate. However, because
(\ref{PAbsFormula}) does not suffer from the numerical 
instabilities that can afflict (\ref{PScatFormula}), we 
are unaware of any situation in which the use of 
(\ref{AlternativePAbsFormula}) over (\ref{PAbsFormula}) is
worth the added computational cost.
%%%%%%%%%%%%%%%%%%%%%%%%%%%%%%%%%%%%%%%%%%%%%%%%%%%%%%%%%%%%%%%%%%%%%%
%%%%%%%%%%%%%%%%%%%%%%%%%%%%%%%%%%%%%%%%%%%%%%%%%%%%%%%%%%%%%%%%%%%%%%
%%%%%%%%%%%%%%%%%%%%%%%%%%%%%%%%%%%%%%%%%%%%%%%%%%%%%%%%%%%%%%%%%%%%%%

%%%%%%%%%%%%%%%%%%%%%%%%%%%%%%%%%%%%%%%%%%%%%%%%%%%%%%%%%%%%%%%%%%%%%%
%%%%%%%%%%%%%%%%%%%%%%%%%%%%%%%%%%%%%%%%%%%%%%%%%%%%%%%%%%%%%%%%%%%%%%
%%%%%%%%%%%%%%%%%%%%%%%%%%%%%%%%%%%%%%%%%%%%%%%%%%%%%%%%%%%%%%%%%%%%%%
\section{Overlap Integrals for RWG Basis Functions}
\label{RWGAppendix}
In the main text we defined various types of overlap integrals
involving pairs of basis functions. Here we present closed-form
expressions for these integrals for the particular choice 
of RWG basis functions~\cite{RWG}. 

The overlap integrals are 
%====================================================================%
\begin{align*}
O^{\times}_{\alpha\beta}
&=\int
  \vb b_\alpha \cdot 
  \Big[\vbhat{n} \times \vb b_\beta(\vb x)\Big]  \, d\vb x 
\\
 O^{\bullet}\IFAB
&\equiv \int 
   \vbhat{n}_i(\vb x) 
   \vb b_\alpha^*(\vb x) \cdot \vb b_\beta(\vb x) 
   d\vb x
\\
 O^{\nabla \nabla}\IFAB
&\equiv \int 
   \vbhat{n}_i(\vb x)
   \big[\nabla \cdot \vb b_\alpha^*(\vb x)\big] 
   \big[\nabla \cdot \vb b_\beta(\vb x)\big]
   d\vb x
\\
 O^{\times \nabla}\IFAB
&\equiv \int 
   \big[\vbhat{n}(\vb x)\times \vb b_\alpha^*(\vb x)\big]_i
   \big[\nabla \cdot \vb b_\beta(\vb x)\big]
   d\vb x
\\
 O^{\bullet}\ITAB
&\equiv \int 
   \Big[ \big(\vb x - \vb x_0\big) \times \vbhat{n}(\vb x) \Big]_i
   \vb b_\alpha^*(\vb x) \cdot \vb b_\beta(\vb x) 
   d\vb x
\\
 O^{\nabla \nabla}\ITAB
&\equiv \int 
   \Big[ \big(\vb x - \vb x_0\big) \times \vbhat{n}(\vb x) \Big]_i
   \big[\nabla \cdot \vb b_\alpha^*(\vb x)\big] 
   \big[\nabla \cdot \vb b_\beta(\vb x)\big]
   d\vb x
\\
 O^{\times \nabla}\ITAB
&\equiv \int 
   \Big\{ \big(\vb x - \vb x_0\big) \times 
          \Big[ \vbhat{n}(\vb x)\times \vb b_\alpha^*(\vb x)\Big]
   \Big\}_i
   \big[\nabla \cdot \vb b_\beta(\vb x)\big]
   d\vb x
\end{align*}
%====================================================================%
where the integrations are over 
$\sup \vb b_\alpha \cap \sup\vb b_\beta.$
%the common support of $\vb b_\alpha,\vb b_\beta.$ 
%####################################################################%
\begin{figure}
\begin{center}
\resizebox{0.5\textwidth}{!}{\includegraphics{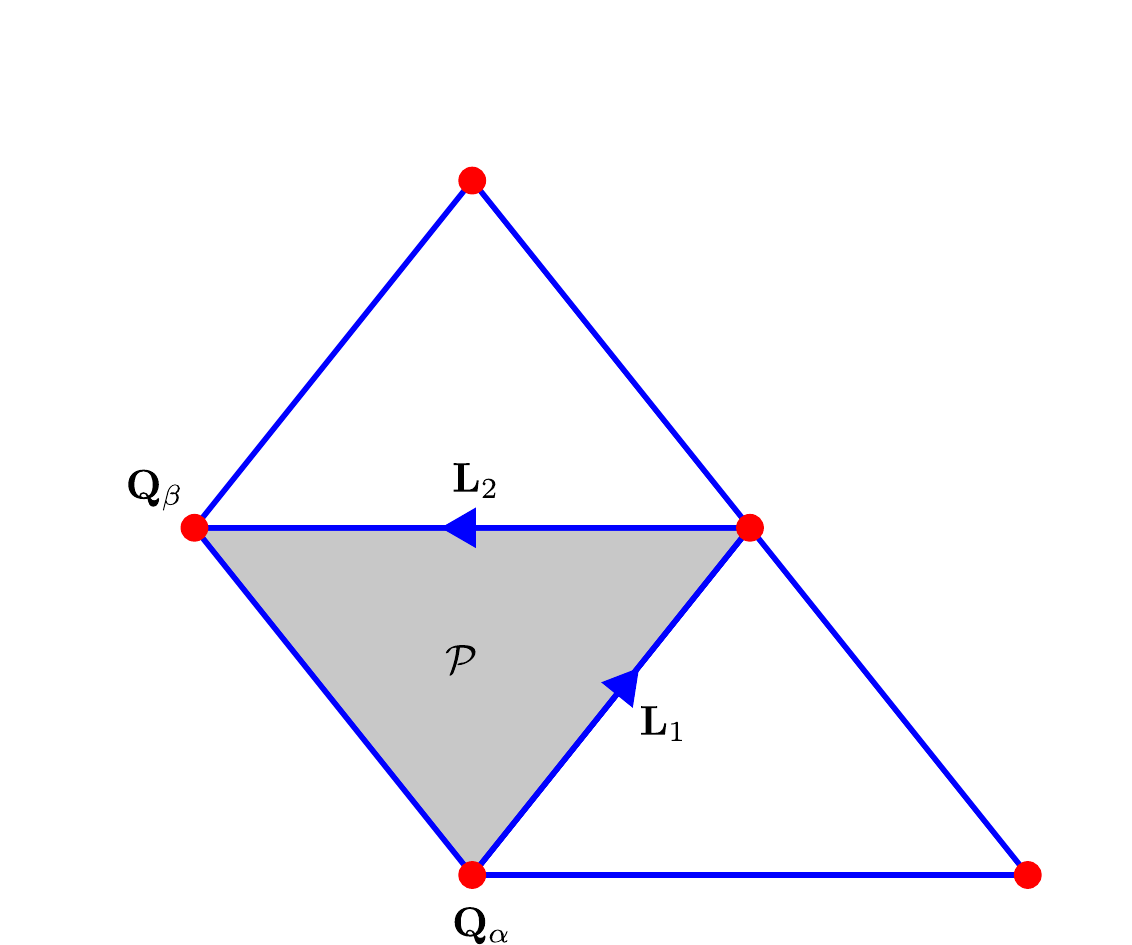}}
\caption{Notation for calculation of overlap integrals between RWG
basis functions. $\mathcal{P}$ is a panel on which two RWG basis
functions $\vb b_\alpha, \vb b_\beta$ are nonvanishing.
$\vb Q_\alpha$ and $\vb Q_\beta$ are the current source/sink
vertices of the two basis functions. Note that the diagonal 
elements of the overlap matrices involve the case 
$\vb b_\alpha=\vb b_\beta$ and hence $\vb Q_\alpha=\vb Q_\beta$.}
\label{RWGFigure}
\end{center}
\end{figure}
%####################################################################%
Each RWG basis function is supported on a 
pair of flat triangular panels, and the overlap integrals have the structure
%====================================================================%
\numeq{OabOp}
{
 O_{\alpha\beta}
=\sum_{\mc P} 
  \frac{\sigma_\alpha \sigma_\beta l_\alpha l_\beta}{2A}
  O_\pan
}
%====================================================================%
where the sum is over panels $\mc P$ common to the supports
of $\vb b_\alpha$ and $\vb b_\beta$ 
(there may be 0, 1, or 2 such panels), 
$A$ is the panel area,
$\sigma_{\alpha,\beta}=\pm$ according as 
$\mc P$ is the positive or negative panel for basis function
$\vb b_{\alpha,\beta},$ and $l_{\alpha,\beta}$ is the
length of the edge associated with $\vb b_{\alpha,\beta}$.
In what follows we give formulas for just the single-panel 
contributions $O_\pan$. 

If we fix the index $\alpha$ in (\ref{OabOp}), then 
the sum on the RHS of that equation 
\textbf{(a)} receives nonvanishing contributions from
two panels if $\beta=\alpha$,
\textbf{(b)} receives nonvanishing contributions from
one panel if $\vb b_\beta$ $(\beta \ne \alpha)$ is one
of the four RWG basis functions that shares a common panel 
of support with $\vb b_\alpha$,
\textbf{(c)} vanishes otherwise. Thus each of the $\mathcal{O}$
matrices has at most $N\subt{NZ}=5$ nonzero entries
per row. ($O^\times$ has $N\subt{NZ}=4$ because the 
diagonal matrix element vanishes in that case.) The force
and torque matrices defined by Equations 
(\ref{ForceBlockMatrix}) and (\ref{TorqueBlockMatrix}) 
have $2N\subt{NZ}$ nonzero entries per row.

Our notational conventions for computing $O_\pan$ 
are summarized in Figure \ref{RWGFigure}, which depicts a single panel 
$\mathcal{P}$
common to the supports of two basis functions
$\vb b_\alpha, \vb b_\beta.$ 
$\vb Q_{\alpha,\beta}$ are the
source/sink vertices for the two basis functions.
We parameterize points in $\pan$
according to
$$ \vb x(u,v) = \vb Q_\alpha + u \vb L_1 + v \vb L_2 $$
in terms of which a general integral takes the form
$$ \int_{\mathcal P} I(\vb x) \, d\vb x
   =2A\int_{0}^1 du \int_0^u dv \, I\big(\vb x(u,v)\big).
$$
The values of the RWG basis functions at $\vb x$ 
are 
\begin{align*}
\vb b_\alpha(\vb x) 
 &= \frac{l_\alpha}{2A}\big(\vb x - \vb Q_\alpha\big)
  = \frac{l_\alpha}{2A}\big(u\vb L_1 + v\vb L_2\big)
\\
\vb b_\beta(\vb x) 
 &= \frac{l_\beta}{2A}\big(\vb x - \vb Q_\beta\big)
  = \frac{l_\beta}{2A}\big(u\vb L_1 + v\vb L_2+ \Delta\vb Q)
\end{align*}
with
$$ \Delta \vb Q=\vb Q_\beta - \vb Q_\alpha.$$
(Note that $\Delta \vb Q=0$ for the case of diagonal
matrix elements $\alpha=\beta.$)

In what follows, $\vbhat{n}$ will denote the outward-pointing
surface normal to $\mc P$, and for a general vector $\vb V$
we will use the shorthand notation $\wt{\vb V}$
to indicate the result of the cross-product with $\vbhat{n}$:
$$ \wt{\vb V} \equiv \vbhat{n} \times \vb V.$$
In principle, some of the results that follow could be 
simplified by using the relation 
$\vbhat{n}=\pm\frac{1}{2A}(\vb L_1\times \vb L_2)$
and noting that $\vb \Delta Q$ may be expressed
as a linear combination of $\vb L_1$ and $\vb L_2$;
however, such simplifications depend on
the precise orientation of the edges and are 
notationally cumbersome.
%====================================================================%
\subsection*{Power overlap integral}
\begin{align}
O^\times_\pan
&=\int_0^1 \, du \, \int_0^u \, dv \, 
  \Big[ u \vb L_1 + v\vb L_2 \Big]
  \cdot
  \Big[ u \wt{\vb L_1} + v\wt{\vb L_2} + \wt{\Delta \vb Q}\Big]
\nn
&=
  \frac{1}{6}
  (2\vb L_1 + \vb L_2) \times \wt{\Delta \vb Q}.
\end{align}

\subsection*{Force overlap integrals}
\begin{align*}
%--------------------------------------------------------------------%
O^\bullet_{i\text{\scriptsize F}\pan}
&=
  \vbhat{n}_i 
  \int_0^1 \, du \, \int_0^u \, dv \, 
  \Big[ u \vb L_1 + v\vb L_2 \Big]
  \cdot
  \Big[ u \vb L_1 + v\vb L_2 + \Delta \vb Q\Big]
\nn
%--------------------------------------------------------------------%
&= \vbhat{n}_i\left[
   \frac{|\vb L_1|^2}{4} + \frac{\vb L_1 \cdot \vb L_2}{4}
   +\frac{\vb L_1 \cdot \Delta \vb Q}{3} 
    \frac{|\vb L_2|^2}{12} + \frac{\vb L_2 \cdot \Delta \vb Q}{6}\right].
\\
%--------------------------------------------------------------------%
%--------------------------------------------------------------------%
O^{\nabla\nabla}_{i\text{\scriptsize F}\pan}
&=
  4\vbhat{n}_i 
  \int_0^1 \, du \, \int_0^u \, dv 
\nn
&= 2\vbhat{n}_i 
\\
%--------------------------------------------------------------------%
%--------------------------------------------------------------------%
O^{\times\nabla}_{i\text{\scriptsize F}\pan}
&=2\vbhat{n}\times
  \int_0^1 \, du \, \int_0^u \, dv
  \Big[u\vb L_1 + v\vb L_2] 
\nn
&=\Big[\frac{2}{3}\wt{\vb L_1} + \frac{1}{3}\wt{\vb L_2}\Big]_i
\end{align*}

\subsection*{Torque overlap integrals}

In what follows we will work in a coordinate system
with origin at $\vb x_0$, the point about which we 
figure the torque.
\begin{align*}
%--------------------------------------------------------------------%
O^\bullet_{i\text{\scriptsize T}\pan}
&=
  -\int_0^1 \, du \, \int_0^u \, dv \, 
  \Big[ \wt{\vb Q_\alpha} + u \wt{\vb L_1} + v\wt{\vb L_2} \Big]_i
  \bigg\{
\\
&\qquad\qquad \Big[ u \vb L_1 + v\vb L_2 \Big]
  \cdot
  \Big[ u \vb L_1 + v\vb L_2 + \Delta \vb Q\Big]\bigg\}
\\
&=\,
 -\Big[\wt{\vb Q_\alpha}\Big]_i O^\bullet_{i\text{\scriptsize T}\pan}
\\
& \quad
  -\Big[\wt{\vb L_1}\Big]_i
                  \left\{   \frac{|\vb L_1|^2}{5}  
                        + \frac{\vb L_1 \cdot \vb L_2}{5}
                        + \frac{\vb L_1\cdot \Delta \vb Q}{4}
                  \right.\\
&\hspace{1.5in}   \left.
                        + \frac{|\vb L_2|^2}{15}
                        + \frac{\vb L_2 \cdot \Delta \vb Q}{8}
                  \right\}
\\
&\quad
  -\Big[\wt{\vb L_2}\Big]_i
                  \left\{   \frac{|\vb L_1|^2}{10} 
                        + \frac{2\vb L_1 \cdot \vb L_2}{15}
                        + \frac{\vb L_1\cdot \Delta \vb Q}{8}
                  \right.\\
&\hspace{1.5in}   \left.
                         + \frac{|\vb L_2|^2}{20}
                         + \frac{\vb L_2 \cdot \Delta \vb Q}{12}
                  \right\}
\\
%--------------------------------------------------------------------%
%--------------------------------------------------------------------%
O^{\nabla\nabla}_{i\text{\scriptsize T}\pan}
&=-4 \int_0^1 \, du \, \int_0^u \, dv 
   \Big[\wt{\vb Q_\alpha} +u \wt{\vb L_1} +v\wt{\vb L_2} \Big]_i
\\
&=-\Big[2\wt{\vb Q_\alpha} + \frac{4}{3}\wt{\vb L_1} + \frac{2}{3}\wt{\vb L_2}\Big]_i
\\
%--------------------------------------------------------------------%
%--------------------------------------------------------------------%
O^{\times\nabla}_{i\text{\scriptsize T}\pan}
&=2\int_0^1 \, du \, \int_0^u \, dv\left\{
   \Big[\vb Q_\alpha + u\vb L_1 + v\vb L_2\Big]
   \right. \\
&\hspace{1.4in} \times
   \left.
   \Big[u\wt{\vb L_1} + v\wt{\vb L_2}\Big]\right\}_i
\\
&=\vbhat{n}_i\left[   \frac{|\vb L_1|^2}{2} 
                   + \frac{\vb L_1\cdot \vb L_2}{2} 
                   + \frac{|\vb L_2|^2}{6}\right]
\\
&\hspace{0.5in}
  +\frac{1}{3}\Big[\vb Q_\alpha \times \left(2\wt{\vb L_1} + \wt{\vb L_2}\right)\Big]_i.
\end{align*}
%%%%%%%%%%%%%%%%%%%%%%%%%%%%%%%%%%%%%%%%%%%%%%%%%%%%%%%%%%%%%%%%%%%%%%
%%%%%%%%%%%%%%%%%%%%%%%%%%%%%%%%%%%%%%%%%%%%%%%%%%%%%%%%%%%%%%%%%%%%%%
%%%%%%%%%%%%%%%%%%%%%%%%%%%%%%%%%%%%%%%%%%%%%%%%%%%%%%%%%%%%%%%%%%%%%%

%%%%%%%%%%%%%%%%%%%%%%%%%%%%%%%%%%%%%%%%%%%%%%%%%%%%%%%%%%%%%%%%%%%%%%%%%%%
%%%%%%%%%%%%%%%%%%%%%%%%%%%%%%%%%%%%%%%%%%%%%%%%%%%%%%%%%%%%%%%%%%%%%%%%%%%
%%%%%%%%%%%%%%%%%%%%%%%%%%%%%%%%%%%%%%%%%%%%%%%%%%%%%%%%%%%%%%%%%%%%%%%%%%%
\bibliographystyle{IEEEtran}


%%%%%%%%%%%%%%%%%%%%%%%%%%%%%%%%%%%%%%%%%%%%%%%%%%%%%%%%%%%%%%%%%%%%%%%%%%%
%%%%%%%%%%%%%%%%%%%%%%%%%%%%%%%%%%%%%%%%%%%%%%%%%%%%%%%%%%%%%%%%%%%%%%%%%%%
%%%%%%%%%%%%%%%%%%%%%%%%%%%%%%%%%%%%%%%%%%%%%%%%%%%%%%%%%%%%%%%%%%%%%%%%%%%

\end{document}